\documentclass{article}
\usepackage{naturetex}
\usepackage{booktabs}
\usepackage{multicol}
\usepackage{multirow}
\usepackage{amsmath}
\usepackage[mathlines]{lineno}
\usepackage{amssymb}
\usepackage{algorithm}
\usepackage[noend]{algpseudocode}
\usepackage{setspace}
\usepackage{caption}
\usepackage{threeparttable}
\usepackage{pdflscape}
\usepackage{array}

\newcolumntype{C}[1]{>{\centering\arraybackslash}p{#1}}

\newcommand{\name}{Frogent }
\newcommand{\nam}{Frogent}
\newcommand{\oa}{Orchestrate agent}
\newcommand{\ra}{Retrieve agent}
\newcommand{\fa}{Forge agent}
\newcommand{\ga}{Gauge agent}
\newcommand{\keywords}[1]{\textbf{Keywords:} #1}

\begin{document}
\maketitle

{\setstretch{1.0}
\section*{Abstract}
Drug discovery is a complex, multi-step pipeline that remains heavily dependent on manual, experience-driven operations; meanwhile, existing customized artificial intelligence tools are fragmented across web applications, desktop software, and code libraries, resulting in incompatible interfaces and inefficient, burdensome workflows. To overcome these challenges, we propose FROGENT, a \textbf{f}ull-p\textbf{ro}cess dru\textbf{g} d\textbf{e}sign multi-age\textbf{nt} system that leverages the planning, reasoning, and tool-use capabilities of large language models (LLMs) to unify drug discovery within a closed-loop and autonomous framework. FROGENT is a collaborative multi-agent system comprising a central Orchestrate Agent for strategic workflow coordination and three distributed agents, Retrieve, Forge, and Gauge, that employ dynamic biochemical databases, extensible tool libraries, and task-specific computational models via the Model Context Protocol. This architecture enables end-to-end execution of complex drug discovery pipelines, covering target identification, small-molecule generation, peptide optimization, and retrosynthetic planning. Across eight benchmarks spanning core drug discovery tasks, FROGENT consistently outperforms six increasingly advanced ReAct-style agents. Case studies further demonstrate its practicality and generalization across real-world small-molecule and peptide design scenarios. Overall, FROGENT not only achieves substantial gains in efficiency and accuracy, but also demonstrates the strong potential of LLM-based agentic systems to autonomously orchestrate drug development pipelines, significantly reducing, or even replacing, reliance on manual, experience-driven human intervention. 

\keywords{Drug Design, Multi-Agent System, Large Language Model, Model Context Protocol.}
}

\section*{Introduction}
Traditional drug development is a capital-intensive and time-intensive endeavor, frequently requiring more than a decade and multi-billion-dollar investments to deliver a single therapeutic to market. Recent advances in artificial intelligence (AI) are catalyzing a paradigm shift in pharmaceutical research by enabling data-driven, scalable, and more efficient discovery strategies \cite{sadybekov2023computational}. AI-driven methodologies are now being systematically integrated across the drug discovery pipeline, encompassing target identification \cite{pun2023ai}, de novo molecular design \cite{zhang2024deep}, and toxicity and safety assessment \cite{wilde2024implementation}. Despite substantial gains in accuracy and cost-effectiveness, the AI ecosystem for drug discovery remains fragmented. Most existing approaches are realized as isolated web services, standalone software, or task-specific code modules, each addressing discrete stages of the pipeline rather than an integrated, end-to-end workflow \cite{zhang2025artificial}. This lack of interoperability and end-to-end automation constrains both operational efficiency and the holistic evaluation of drug candidates.

Fortunately, the rapid evolution of large language models (LLMs), including GPT4 \cite{achiam2023gpt}, Qwen3 \cite{yang2025qwen3}, and DeepSeek-R1 \cite{guo2025deepseek}, has demonstrated strong capabilities in reasoning, planning, decision-making, and tool-augmented action execution within interactive environments. Consequently, LLM-based agents are particularly well suited to addressing complex, multi-step drug design problems, as they can operate effectively in structured scientific settings, support long-horizon reasoning and planning, and systematically decompose the drug discovery process into interdependent subtasks \cite{gao2024empowering}. Moreover, LLM-based multi-agent systems provide a principled paradigm for drug design by harnessing collective intelligence while preserving the functional specialization of individual agents, coordinating communication and collaboration among multiple domain-specific agents. Such systems can integrate heterogeneous expertise and orchestrate tool-augmented workflows, thereby enabling more efficient and autonomous end-to-end drug design \cite{zhao2025llm}.

Considerable efforts have been devoted to the development of AI agents for drug design and discovery. Representative systems span multiple stages of the pipeline. For target discovery, OriGene is a self-evolving multi-agent system that functions as a virtual disease biologist, enabling the systematic identification of mechanistically grounded therapeutic targets \cite{zhang2025origene}. In drug–target interaction modeling, DrugAgent introduces an automated programming framework in which an LLM-based planner generates and refines solution strategies while an instructor module integrates domain knowledge \cite{liu2024drugagent}, and DrugMCTS synergistically combines retrieval-augmented generation, multi-agent collaboration, and Monte Carlo tree search for drug repositioning \cite{yang2025drugmcts}. For lead identification and optimization, ChatDrug integrates prompting, retrieval, domain feedback, and conversational modules to streamline molecular editing tasks \cite{liu2024conversational}. Beyond in silico design, VirtualLab formulates an LLM-based principal investigator agent that collaborates with human researchers to design SARS-CoV-2 nanobodies \cite{swanson2025virtual}. In the context of chemical synthesis, LLM-RDF employs six specialized LLM-based agents to guide the end-to-end synthesis development process \cite{ruan2024automatic}, while ChemCrow combines chain-of-thought reasoning with expert-curated chemistry tools to automate organic catalyst synthesis \cite{m2024augmenting}. Nevertheless, most existing agent-based systems are confined to specific stages of the drug design pipeline, thus remain at the level of module-level intelligence augmentation, rather than constituting end-to-end drug design systems. Although DrugPilot \cite{li2025drugpilot} and PharmAgents \cite{gao2025pharmagents} aim to streamline and automate the pharmaceutical research pipeline, they rely on predefined workflows or manually specified agent roles, resulting in largely fixed execution paths that limit the effective exploitation of LLMs’ knowledge-driven reasoning, long-horizon planning, and intermediate reflection capabilities required for truly autonomous, end-to-end drug discovery workflows.

To address these challenges, we introduce \name, a \textbf{f}ull-p\textbf{ro}cess dru\textbf{g} d\textbf{e}sign multi-age\textbf{nt} system that enables end-to-end automation of drug discovery workflows. By explicitly modeling reasoning and planning, \name supports long-horizon, multi-step decision-making, thereby fully leveraging the cognitive capabilities of large language models. In addition, \name adopts a deeply integrated multi-agent collaboration framework that enforces goal alignment, facilitates information sharing, and coordinates strategy optimization across agents, ensuring globally consistent design decisions. Through the integration of language-based interaction with adaptive orchestration, \name lowers technical barriers and streamlines decision-making across the entire pipeline. In contrast to existing agent-based approaches that optimize isolated stages, \name represents a paradigm-level advance toward autonomous drug design. It is designed to support the full discovery process, encompassing problem formulation, objective specification, molecular design, and systematic evaluation. We validate the effectiveness of \name through a comprehensive empirical evaluation. Across eight diverse benchmarks spanning the full drug discovery workflow, \name consistently and substantially outperforms six state-of-the-art baseline methods. Beyond quantitative benchmarking, three representative case studies further demonstrate its practical utility and generalizability across therapeutic modalities. Collectively, these results show that the agentic architecture and specialized tool ecosystem of \name enable efficient and effective navigation of the inherent complexity of drug discovery. To our knowledge, \name is the first multi-agent drug design framework that integrates and automates end-to-end discovery workflows for both small-molecule and peptide therapeutics, marking a concrete step toward truly autonomous drug discovery beyond semi-automated toolchains.

\section*{Results}

\subsection*{Architecture Design of Frogent}
\begin{figure}[!h] 
\centering \includegraphics[width=1.0\textwidth]{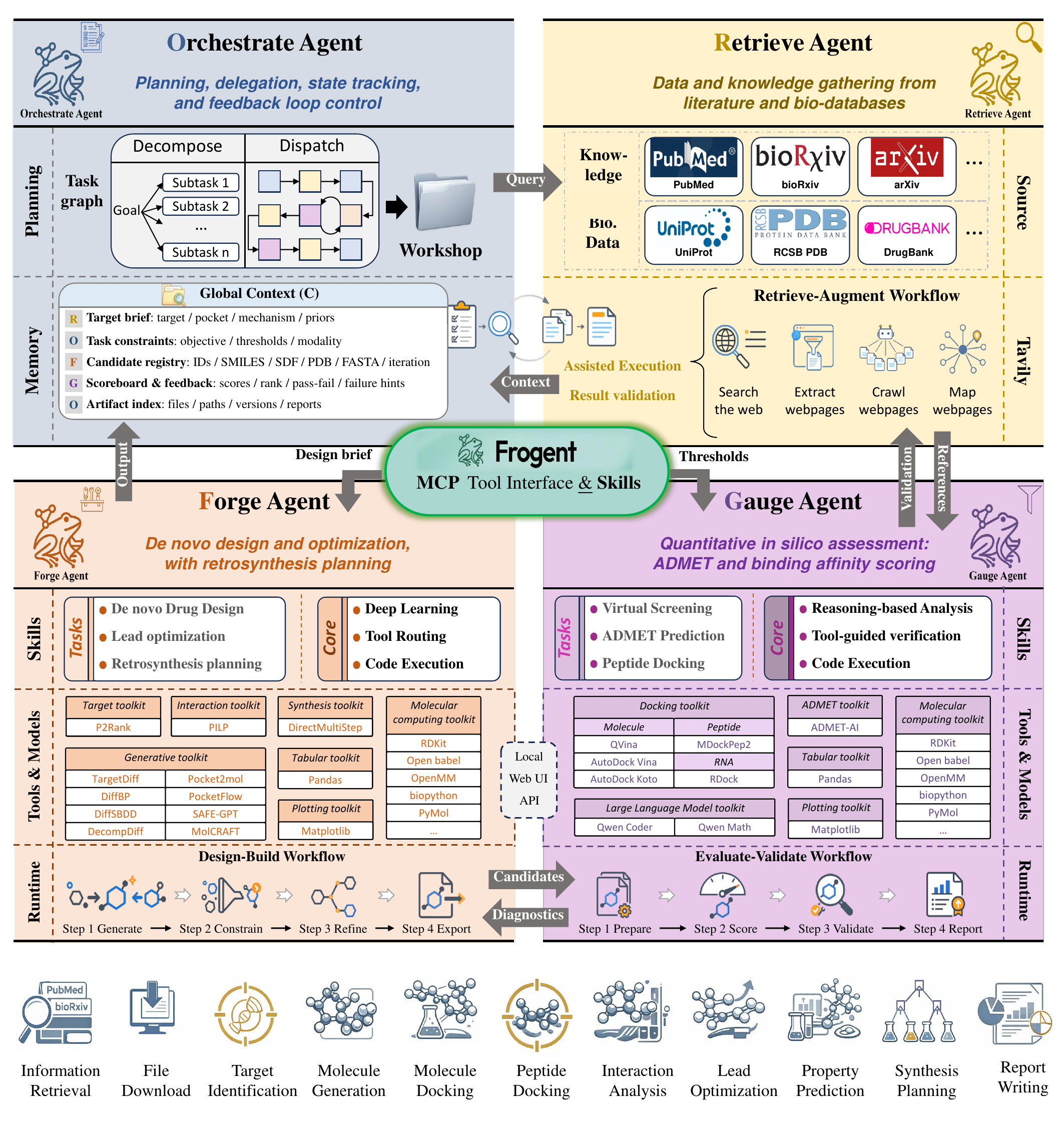} 
\caption{\textbf{Architectural Overview of the \name Multi-Agent System.} 
The system comprises four distinct agents coordinated by a central \oa \ that handles planning, delegation, and feedback loop control. The \ra \ gathers essential data from literature and biological databases. The \fa \ is responsible for all generative tasks, including de novo molecular design, optimization, and retrosynthesis planning. The \ga \ performs quantitative validation for both small molecules and peptides. The bottom panel illustrates the comprehensive suite of 11 drug discovery capabilities that \name can execute autonomously.} 
\label{fig:Architecture} 
\end{figure}

To systematically address the complexity of the end-to-end drug discovery process, we architected \name as a decentralized, collaborative multi-agent system. This design paradigm moves beyond a monolithic agent, embracing a division of labor that assigns distinct responsibilities to specialized agents, as illustrated in Figure \ref{fig:Architecture}. The framework is composed of four cornerstone agents: \oa, \ra, \fa \ and \ga. Upon receiving a user's high-level goal, the \oa, serving as the central controller, initiates the workflow by performing planning: it decomposes the goal into a structured task graph of executable subtasks and dispatches them to the appropriate specialized agents. A central feature is the Global Context, a dynamic memory module managed by the \oa, which maintains the state of the entire workflow. This context serves as a shared workspace, registering critical artifacts from each agent's operations, such as target information from the \ra, user-defined task constraints, candidate molecules from the \fa, and evaluation scores from the \ga.

The \ra functions as the framework's epistemic module, responsible for foundational capabilities such as Information Retrieval and File Download. In response to queries from the \oa \ or other peer agents, it performs data and knowledge gathering by leveraging its access to both static knowledge bases and a dynamic Retrieve-Augment Workflow. This workflow enhances its capabilities with real-time web searching, webpage extraction, and content mapping to incorporate emergent information, which is then used for both task execution and result validation. A key application of this agent is in Target Identification, where it performs iterative searches to identify and prioritize disease-relevant targets based on scientific literature and biological databases.

The \fa \ acts as the framework's creative core, executing its primary tasks, Molecule Generation, Lead Optimization, and Synthesis Planning, through a synergy of Deep Learning, Tool Routing, and Code Execution. To achieve this, its action space is equipped with an extensive toolkit, including protein pocket discovery tools, protein-ligand interaction analyzers, a suite of generative models, retrosynthesis planners, and a rich library of molecular computing and visualization tools.

Complementing the generative capabilities is the \ga, which serves as the critical validation and filtering module. It is responsible for all quantitative assessments, including Molecule Docking, Peptide Docking, Property Prediction, and Interaction Analysis by leveraging a combination of Reasoning-based Analysis, Code Execution, and Tool-guided verification. Its toolkit is similarly comprehensive, featuring specialized docking software for both small molecules and peptides, ADMET prediction models, and various computational and LLM-based tools for analysis. A key innovation within \name \ is the closed-loop optimization cycle formed by the dynamic interplay between the \fa \ and \ga. Within this loop, the \fa \ is responsible for generating and refining molecules, which are then passed to the \ga \ for rigorous scoring and validation. This iterative ``design-evaluate-refine" process, supervised by the \oa, continues for multiple rounds until candidates that satisfy the initial constraints are produced.

\begin{figure}[htbp] 
\centering \includegraphics[width=1\textwidth]{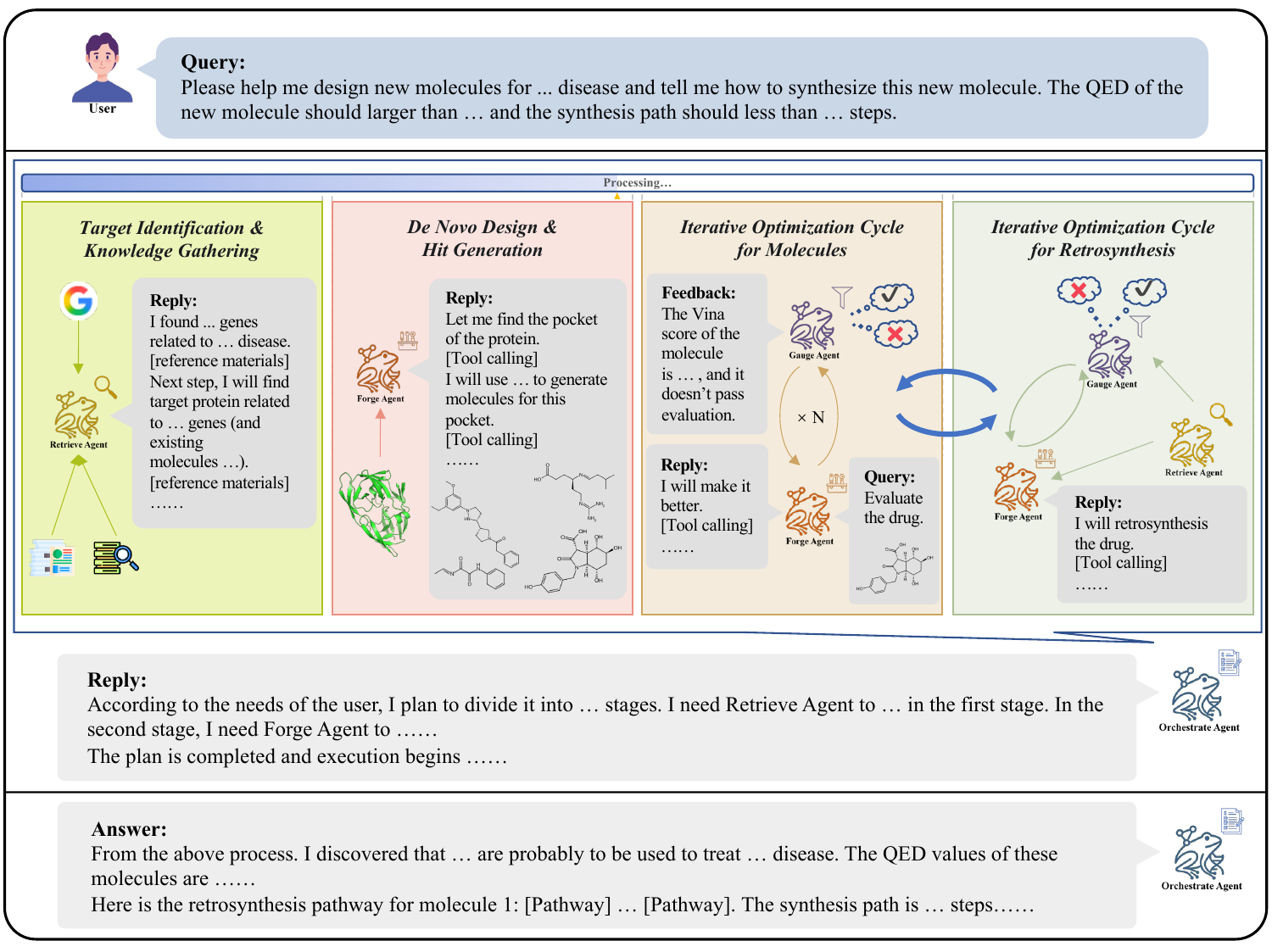} 
\caption{\textbf{Conceptual workflow of \nam.} 
The \oa \ coordinates a multi-stage drug discovery campaign initiated by a single user query. The workflow progresses from initial knowledge gathering and hit generation to the core iterative optimization cycles, where collaboration between the \fa \ and \ga \ refines both molecular properties and synthetic feasibility before the \oa \ synthesizes a final solution.}
\label{fig:overview} 
\end{figure}

Coordinated by the \oa, this team of specialists works in concert to navigate the drug discovery pipeline autonomously, as illustrated in the conceptual workflow depicted in Figure~\ref{fig:overview}. Upon receiving a user's high-level goal, the \oa \ first analyzes the request and decomposes it into a multi-stage strategic plan. For each stage, it delegates a primary task to a specific specialized agent responsible for that domain. Within each stage, an agent can accomplish its objectives in two ways: either by autonomously utilizing its own suite of integrated tools and knowledge bases, or by engaging in multi-turn, peer-to-peer interactions with other specialized agents to resolve sub-task dependencies. Upon the completion of each stage, the \oa \ synthesizes the intermediate results, performs an analysis, and makes a decision on initiating the subsequent stage of the plan. After all predefined stages have been executed, the \oa \ aggregates and analyzes the complete set of findings from the entire workflow, generating a comprehensive final report that is delivered to the user. 

\subsection*{Evaluation}

To comprehensively evaluate the capabilities of \nam, we benchmarked it against six progressively capable baseline agent architectures: (1) a base LLM (GPT-4o) without access to any tools, representing a strong zero-tooling baseline; (2) a base LLM (Qwen3-32B) under the same zero-tooling setting, serving as a representative open-source foundation model; (3) a standard ReAct agent that uses chain-of-thought reasoning to make function calls; (4) ReAct+Code, which augments the agent with a Python code interpreter for computational tasks; (5) ReAct+Literature, which provides access to unstructured knowledge via PubMed and ArXiv; and (6) ReAct+Code+Literature, the most powerful baseline combining both computational and knowledge retrieval tools. Against this competitive landscape, all agents were assessed across a suite of eight benchmarks. These benchmarks are designed to span the full drug discovery pipeline, from foundational knowledge retrieval to real-world generative tasks.

\begin{figure}[htbp] 
\centering \includegraphics[width=1.0\textwidth]{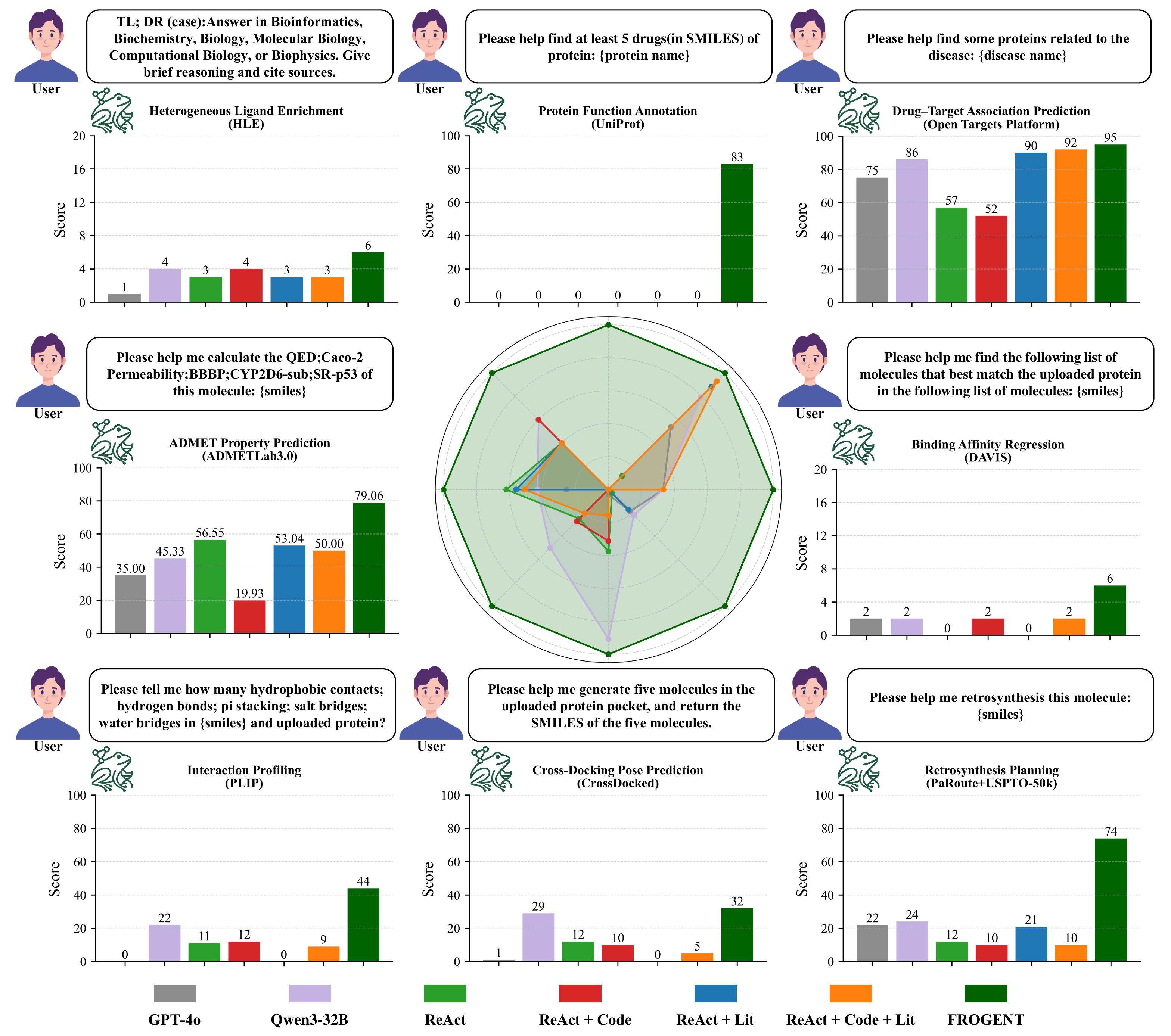} 
\caption{\textbf{Overall Performance of Each Agent on Benchmarks.} 
Performance evaluation of \name  against six baseline models across eight diverse benchmarks spanning the full drug discovery workflow. The evaluated tasks span the entire workflow, from foundational knowledge assessment and structured data retrieval to critical hit-to-lead activities and mechanistic analysis. The evaluation culminates with complex, end-stage challenges and ensuring synthetic feasibility. \name  consistently outperforms all baselines, highlighting the effectiveness of its integrated architecture and specialized tool suite in navigating the complexities of drug discovery.} 
\label{fig:evaluations} 
\end{figure}

In these evaluations, \name  demonstrated superior performance across the board, as illustrated in Figure  \ref{fig:evaluations}. The evaluation began by testing foundational biomedical knowledge on a benchmark derived from the Humanity's Last Exam (HLE) \cite{phan2025humanity}, a task crucial for ensuring an agent can interpret complex biological queries accurately. \name  achieved a score of 6, significantly outperforming the best baseline. The modest performance of the base LLMs on this task was expected, as it relies on their internal knowledge. However, this superiority was starkly evident in tasks requiring agents to reason over complex structured data. When tasked with retrieving known active molecules for given proteins from UniProt \cite{uniprot2025uniprot}, which is a common starting point for lead optimization campaigns. \name  achieved a near-perfect score of 83, while all baseline agents completely failed. This result highlights a critical limitation of general-purpose agents. Without the domain-specific knowledge of how to query and parse the outputs from specialized scientific databases and interfaces, they are incapable of navigating the complex schemas of essential scientific repositories. Similarly, in a disease-target validation task using data from the Open Targets Platform \cite{buniello2025open}, which is necessary for de-risking the entire project by confirming the therapeutic hypothesis, \name  scored 95, surpassing the strongest baseline and demonstrating its robust data retrieval capabilities. 

In core computational chemistry workflows, \nam's integration of specialized tools provided a decisive advantage. To minimize losses in the drug discovery process, it is crucial to conduct early filtration tests on small molecules based on their ADMET properties. We evaluated agents on a property prediction task for five key endpoints, using test data sourced from ADMETLab 3.0 \cite{fu2024admetlab}. \name  scored 79.06, demonstrating a significant improvement over the top-performing baseline. On a virtual screening benchmark constructed from the DAVIS dataset \cite{davis2011comprehensive}, which simulates the critical hit-finding process of identifying the best potential drug from a large pool of candidates, \name  achieved a score of 6, tripling the performance of the best baseline. To assess its ability to elucidate molecular mechanisms of binding, a vital step that informs rational lead optimization, we evaluated agents on an interaction profiling benchmark using ground-truth data generated by PLIP \cite{salentin2015plip}. In this task, agents had to correctly identify key protein-ligand interactions, \name  doubled the score of the nearest competitor, showcasing the value of its expert tools over generic code execution.

Finally, we tested these agents on the most challenging generative and planning tasks. In a de novo design task using protein pockets from the CrossDocked dataset \cite{francoeur20203d}, a test of its creative capacity to invent novel molecules, it generated candidates with superior predicted properties, achieving a leading score of 32, which is significantly higher than the best baseline. For retrosynthesis planning against benchmarks from USPTO-50k and PaRoute \cite{genheden2022paroutes}, which serves as the ultimate reality check to ensure a designed molecule is synthetically feasible, it achieved a high score of 74, more than tripling the performance of the best baseline. This particular result underscores the core value of the specialized models. Although the baselines are equipped with a code interpreter, they cannot replicate the performance of specialized pretraining models. The consistently superior performance across this wide spectrum of benchmarks validates that simply equipping an LLM with generic tools is insufficient. Instead, \nam's robust and versatile architecture, with its curated data, specialized tools, and expert AI models, is uniquely capable of autonomously navigating the end-to-end drug discovery process.

\subsection*{Case Studies}

To demonstrate the versatility and practical effectiveness of the \nam, we present three representative case studies. Each highlights key stages in the drug discovery pipeline and illustrates how the platform can autonomously manage complex scientific tasks. These cases go beyond simple integration of data, tools, and AI models, showcasing the dynamic interplay among agents as they reason, plan, and execute a coherent pipeline.

\subsubsection*{End-to-End Drug Design for Cardiomegaly-congestive Heart Failure}
\label{case: 1}
Target discovery forms the foundation of modern pharmaceutical research. However, it remains a leading cause of failure in late-stage clinical trials due to inadequate validation or selection of non-actionable targets \cite{waring2015analysis, hughes2011principles}. A pipeline that couples target discovery with downstream design tasks can address this challenge by aligning molecular generation efforts with validated mechanisms from the outset \cite{mak2024artificial}.

\begin{figure}[t] 
\centering \includegraphics[width=1.0\textwidth]{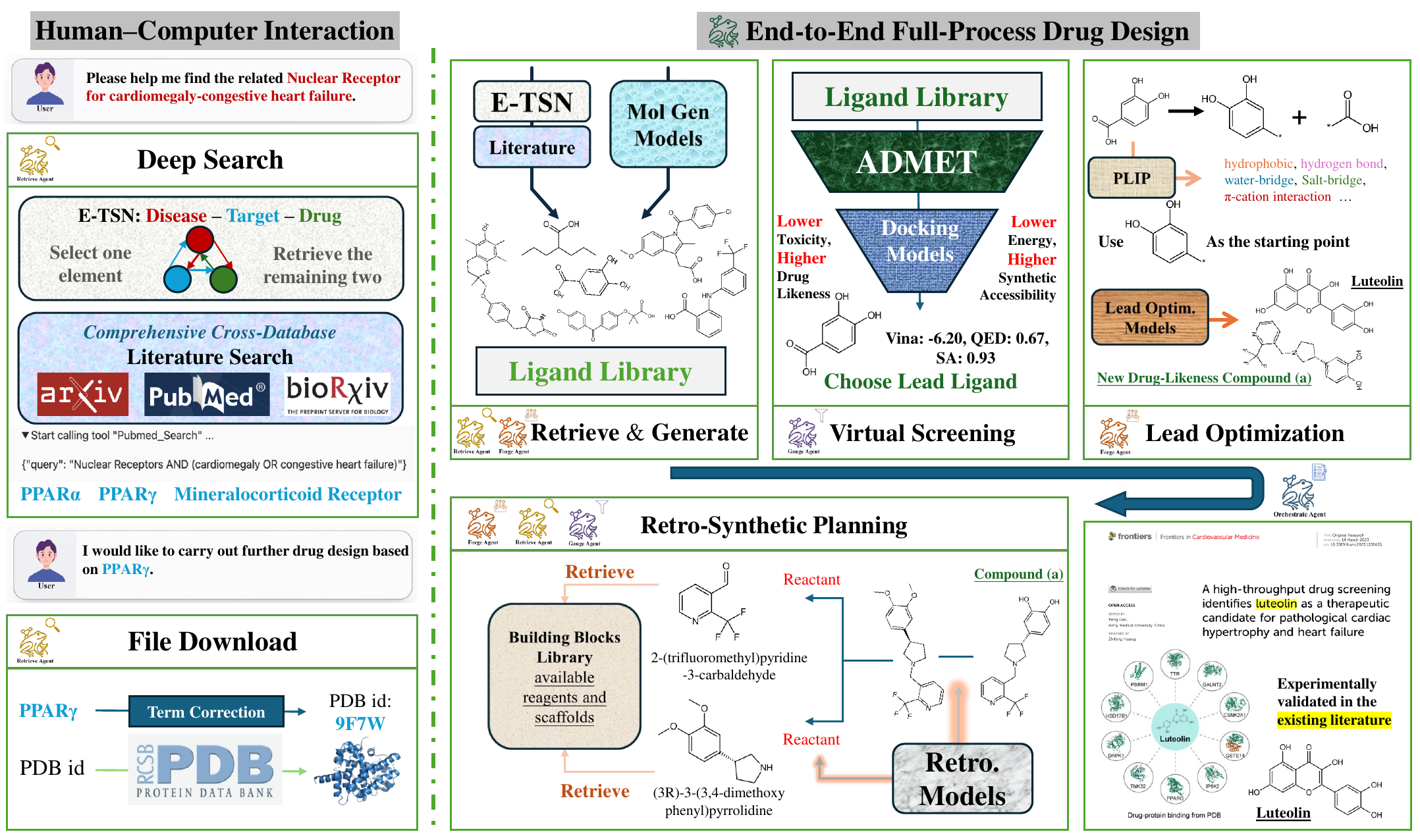} 
\caption{\textbf{End-to-End Drug Design for Cardiomegaly-congestive Heart Failure.} 
Following a user query, the \oa \  tasks the \ra \ to identify PPAR$\gamma$ as a relevant target, which provides the necessary context for the \fa \ to generate initial drug candidates. An iterative optimization cycle driven by collaboration between the \fa \ and \ga \ refines these candidates. The multi-agent process successfully rediscovers the known active molecule Luteolin and designs a novel, superior candidate, Compound (a), complete with a viable retrosynthesis plan.} 
\label{fig:pipeline} 
\end{figure}

To evaluate this collaborative capability, we initiated a comprehensive discovery cycle for cardiomegaly and congestive heart failure, as depicted in Figure \ref{fig:pipeline}. The process began when the \oa \ received the user's broad query. It first tasked the \ra \ to perform a comprehensive literature and database search, which identified PPAR$\alpha$, PPAR$\gamma$, and Mineralocorticoid Receptor as relevant nuclear receptor targets. Subsequently, guided by the user's prioritization of PPAR$\gamma$ as the target receptor  for the entire design process, the \oa \ directed the \ra \ to secure a high-resolution crystal structure (PDB ID: 9F7W). 

With the target validated and its structure obtained, the \oa \ formulated a design plan and tasked the \fa \ to generate candidate molecules. The \fa \ directed the  \ra \ to gather known ligands from DrugBank to establish a baseline, and employed its own suite of de novo generative models to create novel scaffolds tailored to the protein's binding pocket simultaneously. The resulting library of diverse candidate molecules was then passed by the \oa \ to the \ga \ for rigorous filtering. The \ga \ performed ADMET predictions and molecular docking to assess drug-likeness and binding affinity, returning a ranked list of promising hits. The \oa \ then initiated the critical lead optimization cycle: it fed the evaluation results back to the \fa, which performed fragment-level modifications on the highest-potential scaffolds. This iterative \fa \ loop continued until the candidate's predicted properties met the desired criteria.

The system successfully rediscovered Luteolin, a natural flavonoid with experimentally validated efficacy against pathological cardiac hypertrophy \cite{wang2023high}, underscoring the platform's ability to identify biologically relevant molecules. In parallel, it designed a novel candidate, Compound (a), which surpassed Luteolin in key metrics. Finally, the \oa \ tasked the \fa \ to confirm synthetic feasibility. The \fa \ devised a plausible retrosynthesis plan for Compound (a) by querying the \ga \ and \ra \ for commercially available building blocks, thus completing the computational workflow from a high-level goal to a viable drug candidate.

\subsubsection*{Lead Optimization for Glucagon-Like Peptide 1 Receptor}
\label{case: 2}
Peptide therapeutics represent a rapidly growing class of drugs, but issues of binding affinity, stability, and selectivity often challenge their development \cite{kumar2023recent}. The optimization of a native peptide sequence to enhance its interaction with a target receptor is a critical and complex task in modern drug discovery \cite{lee2019comprehensive}.

\begin{figure}[t] 
\centering \includegraphics[width=1.0\textwidth]{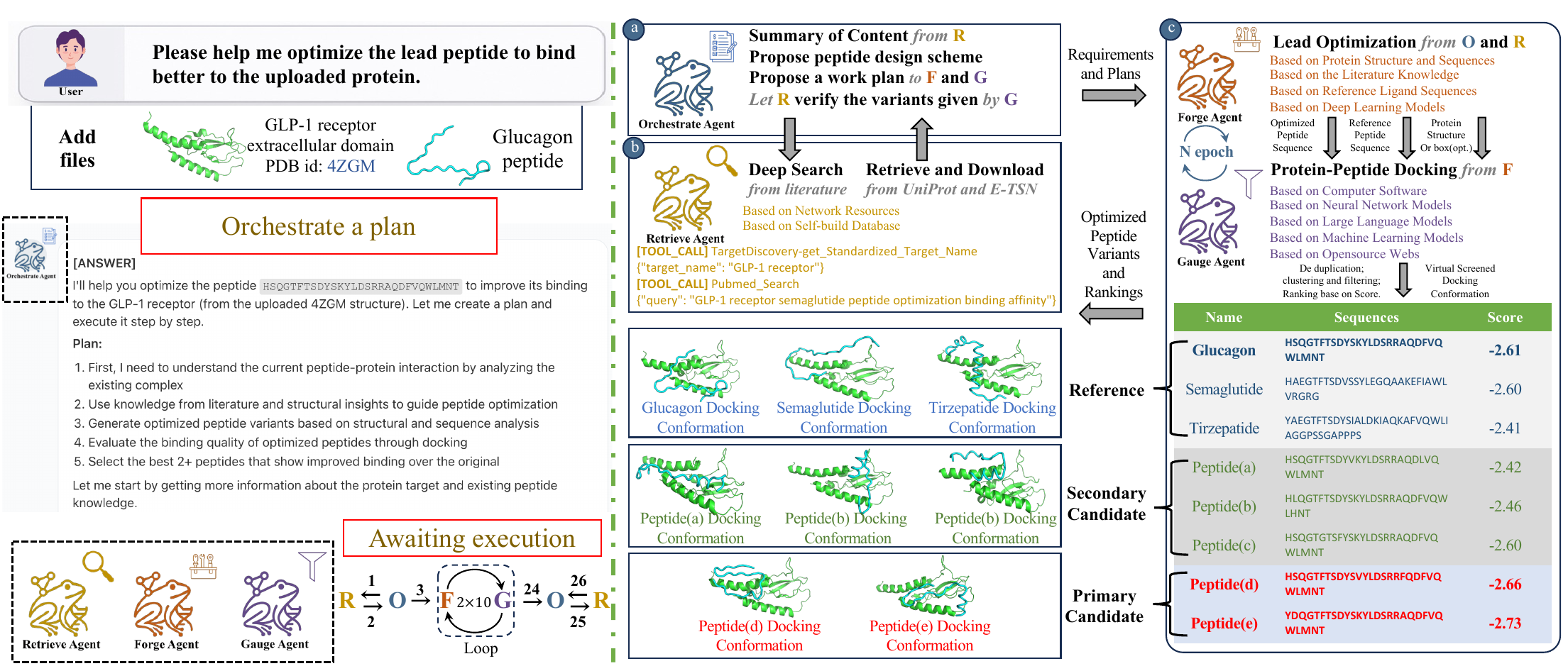} 
\caption{\textbf{Lead Optimization for Glucagon-Like Peptide 1 Receptor.} 
The \oa \ initiates the campaign by tasking the \ra \ to acquire the GLP1R structure (PDB: 4ZGM) and the Glucagon sequence. It supervises a collaborative optimization cycle. The \fa \ generates new peptide variants, and the \ga \ evaluates these peptide variants by predicting their binding affinity. Feedback from this evaluation guides the \fa’s subsequent design rounds, culminating in two novel peptide sequences with predicted binding scores superior to known potent agonists.} 
\label{fig:LeadOptimizationsPep} 
\end{figure}

To demonstrate \nam's versatility beyond small molecules, we tasked it with optimizing the native peptide Glucagon to improve its binding affinity for the Glucagon-like peptide 1 receptor (GLP1R), a key target in type 2 diabetes and obesity. The process, illustrated in Figure \ref{fig:LeadOptimizationsPep}, highlights the framework's ability to handle the unique complexities of peptide design. The \oa \ initiated the task by directing the \ra \ to fetch the 3D structures of the GLP1R protein (PDB ID: 4ZGM) and the amino acid sequence of Glucagon. With this structural context, the \oa \ formulated a peptide optimization plan and tasked the \fa \ to generate a library of promising peptide analogues  via targeted in silico mutagenesis at key positions within the Glucagon sequence.

The core of this case study was the iterative optimization loop tailored for biologics. The \fa \ systematically generated new peptide variants and passed them to the \ga \ for evaluation. The \ga, in turn, utilized the specialized tool to predict the binding affinity and docked conformation of each new peptide to the GLP1R. The resulting docking scores were returned as feedback to the \fa. With this information, the \fa \ can focus on substitutions that showed the most promise and modify them in the next round. After several cycles of this collaborative "design-evaluate-refine" loop, the system converged on two novel peptide sequences. These sequences exhibited significantly more favorable predicted binding scores(-2.73 and -2.66) compared not only to the native Glucagon(-2.61) but also to the known GLP1R agonists, Semaglutide(-2.6) and Tirzepatide(-2.42), under the same in silico conditions. This case study demonstrates \nam's capacity for intelligent, multi-agent-driven optimization of complex biologics, showcasing its adaptability and effectiveness beyond the domain of traditional small molecules.

\subsubsection*{Lead Optimization for Carbonic Anhydrase Inhibitors}
\label{case: 3}
Lead optimization is a resource-intensive phase that transforms initial hit compounds into drug candidates with balanced potency, selectivity, and drug-like properties \cite{wager2010defining, gleeson2011probing}. This stage often demands iterative structure refinement guided by a complex interplay of biological and chemical considerations.

\begin{figure}[t] 
\centering \includegraphics[width=1.0\textwidth]{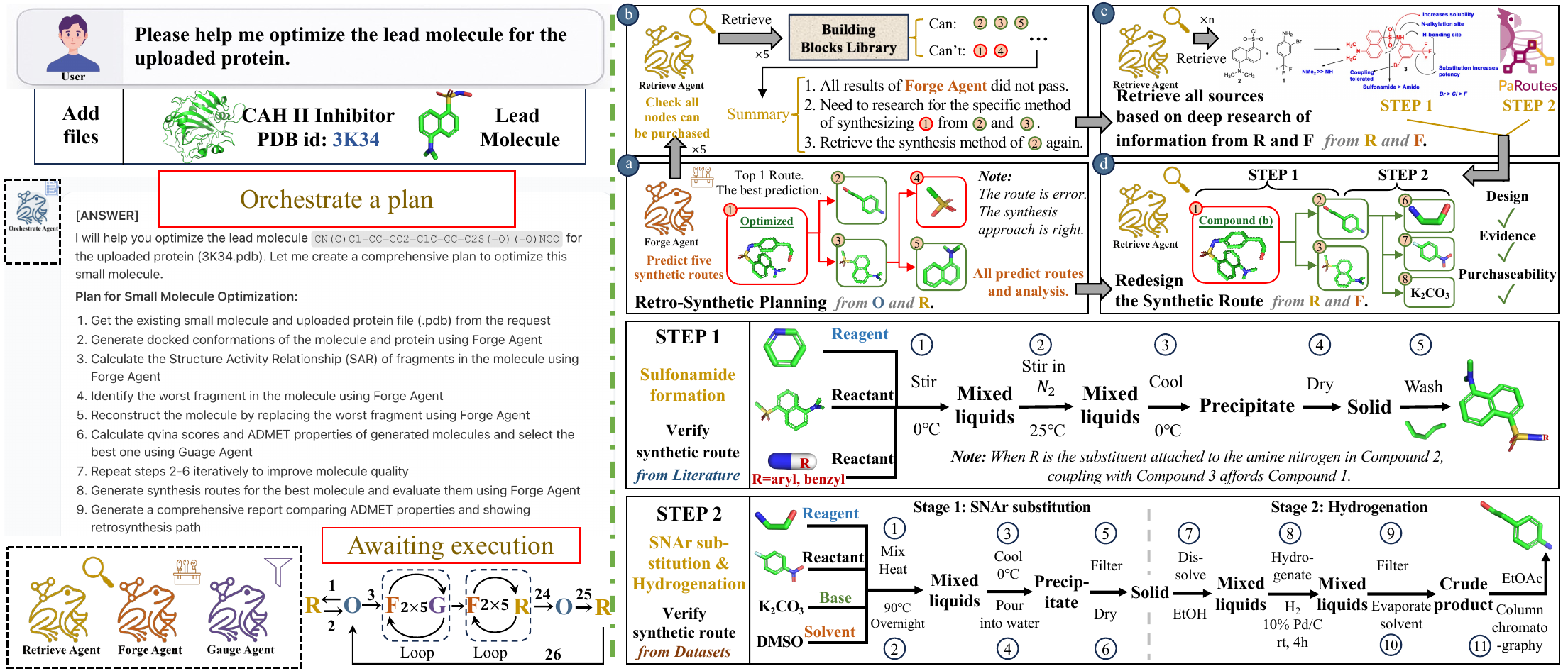} 
\caption{\textbf{Lead Optimization for Carbonic Anhydrase Inhibitors.} 
Starting with a known lead molecule, the agents work collaboratively to analyze, generate, and evaluate a library of analogues. The process yields an improved candidate, Compound (b), with enhanced predicted properties. When an initial retrosynthesis plan generated by the \fa \ proved unfeasible due to non-commercially available precursors, the \ra \ assisted by providing alternative synthesis information. This collaborative repair process resulted in two practical synthesis plans that align with known experimental routes, demonstrating the system's robustness.} 
\label{fig:LeadOptimizationsMol} 
\end{figure}

We assessed \nam's collaborative capability in this context by providing the \oa \ with a known N-substituted sulfonamide inhibitor targeting human Carbonic Anhydrase II (CAH II, PDB ID: 3K34), as shown in Figure \ref{fig:LeadOptimizationsMol}. The \oa \ first tasked the \fa \ to perform a detailed structural analysis, identifying key binding features and solvent-exposed regions suitable for modification. Based on this analysis, the \oa \ initiated the core optimization loop. It directed the \fa \ to use its fragment-based generative models to propose a focused library of new analogues. This library was then passed to the \ga \ for comprehensive evaluation, including virtual screening, ADMET assessment, and redocking.

The \ga \ returned a ranked list of candidates to the \oa, which identified Compound (b) as the top-performing molecule with an improved predicted binding affinity and a more favorable ADMET profile. To complete the campaign, the \oa \ directed the \fa \ to assess synthetic accessibility. The \fa \ initially employed the DirectMultiStep tool, which proposed a chemically valid but strategically flawed retrosynthesis plan. The flaw was identified when the \fa \ tasked the \ga \ and \ra \ to verify the commercial availability of the proposed starting materials; the check failed, as the precursors were found to be non-purchasable.

Recognizing the failure, the \oa \ initiated a ``synthesis repair" sub-task. It directed the \ra \ to search for alternative, documented synthesis methods for similar structural motifs within the scientific literature and the PaRoutes dataset \cite{genheden2022paroutes}. Armed with this new context, the \oa \ then re-tasked the \fa \ to repair the initial pathway. The \fa, now guided by this additional domain knowledge, successfully devised two practical synthesis routes from commercially available starting materials. Notably, this final, repaired synthetic pathway closely matched a recently published and experimentally validated route \cite{babalola2024design}, confirming the platform's ability to not only design molecules but also to intelligently solve complex synthesis challenges by grounding its plans in real-world chemistry through multi-agent synergy.

\section*{Discussion}

\subsection*{LLM-driven Autonomous Agents}
The emergence of LLM-driven autonomous agents represents a significant paradigm shift, extending the capabilities of Large Language Models(LLMs) beyond static text generation to dynamic, goal-oriented problem-solving. At their core, these agents are architected to perform complex reasoning, formulate multi-step plans, and interact with external environments through tool use. Pioneering frameworks such as ReAct \cite{yao2023react} established a powerful methodology by prompting LLMs to interleave reasoning "thoughts" with concrete "actions", creating a synergistic and interpretable loop of planning and execution. Works like Toolformer \cite{schick2023toolformer} further solidified this fundamental ability to interact with the external world. Gorilla \cite{patil2024gorilla} demonstrated that LLMs can be effectively trained to master the use of external APIs, grounding their abstract reasoning in executable, real-world functions. Building upon these single-agent foundations, the field has progressed toward multi-agent systems (MAS), where collaboration is key to solving even more complex challenges. Frameworks like AgentVerse \cite{chen2023agentverse} and MetaGPT \cite{hong2023metagpt} showcase how intricate tasks can be decomposed and assigned to a cohort of specialized agents that collaborate, much like a human team, to achieve a common goal. However, while these general-purpose agentic frameworks provide a powerful blueprint for autonomous systems, they inherently lack the deep domain knowledge and fine-grained, validated toolsets essential for navigating the complexities of specialized scientific domains such as drug discovery. Their effectiveness is contingent on the availability of generic tools, which fall short of the requirements for high-stakes, precision-oriented scientific research.

\subsection*{Agents in Drug Discovery}
The transformative potential of AI has spurred the development of specialized agents tailored for various facets of drug discovery, each addressing specific and complex challenges within the pipeline. For instance, ChemCrow \cite{bran2023chemcrow} demonstrated significant capabilities by augmenting an LLM with a suite of chemistry-focused tools, enabling it to plan and execute tasks in organic synthesis and drug design effectively. In a parallel effort, DrugAgent \cite{liu2024drugagent} focuses on a different yet equally critical bottleneck: automating the machine learning (ML) programming required for drug discovery tasks, acting as an agent that can write, debug, and execute experimental code. In a more interactive paradigm, ChatMol Copilot \cite{sun2024chatmol} serves as a user-centric assistant, translating natural language instructions into complex molecular modeling and computational commands, thereby lowering the barrier for researchers to access sophisticated simulation tools.

Although these systems represent significant advancements, they are characteristically task-oriented. ChemCrow excels in chemical synthesis and DrugAgent in ML experimentation, but neither is designed to autonomously manage the entire workflow, including crucial preceding stages like novel disease target discovery. Similarly, systems like ChatMol Copilot, while powerful, operate primarily as ``copilots" that require continuous human guidance to navigate from one discrete task to the next. Consequently, a critical gap persists: there is no single, unified framework capable of autonomously planning and executing the entire end-to-end drug discovery workflow, from initial biological target discovery to the final design of a retrosynthesis pathway. Existing solutions function as either expert systems for isolated stages or as interactive assistants for specific computations, rather than as truly autonomous, full-process discovery engines. This fragmentation highlights the clear need for a holistic agent like \nam, which is designed to bridge these silos and manage the discovery process from inception to completion.

\subsection*{Limitations}
While \name demonstrates a significant advance in automating the drug discovery workflow, its efficacy is intrinsically linked to the capabilities of its foundational components. The system's reasoning and scientific knowledge are fundamentally bounded by the fidelity of its core LLM, which can still exhibit hallucinations or flawed logic when faced with complex scientific problems. While our multi-agent architecture is designed to mitigate such errors by grounding actions in tool-based observations, the system's ability to recover from a flawed plan is ultimately dependent on the LLM's reasoning fidelity. Looking forward, the continuous evolution of next-generation foundation models will be crucial for enhancing the system's robustness and scientific accuracy.

\begin{figure}[!h] 
\centering \includegraphics[width=0.8\textwidth]{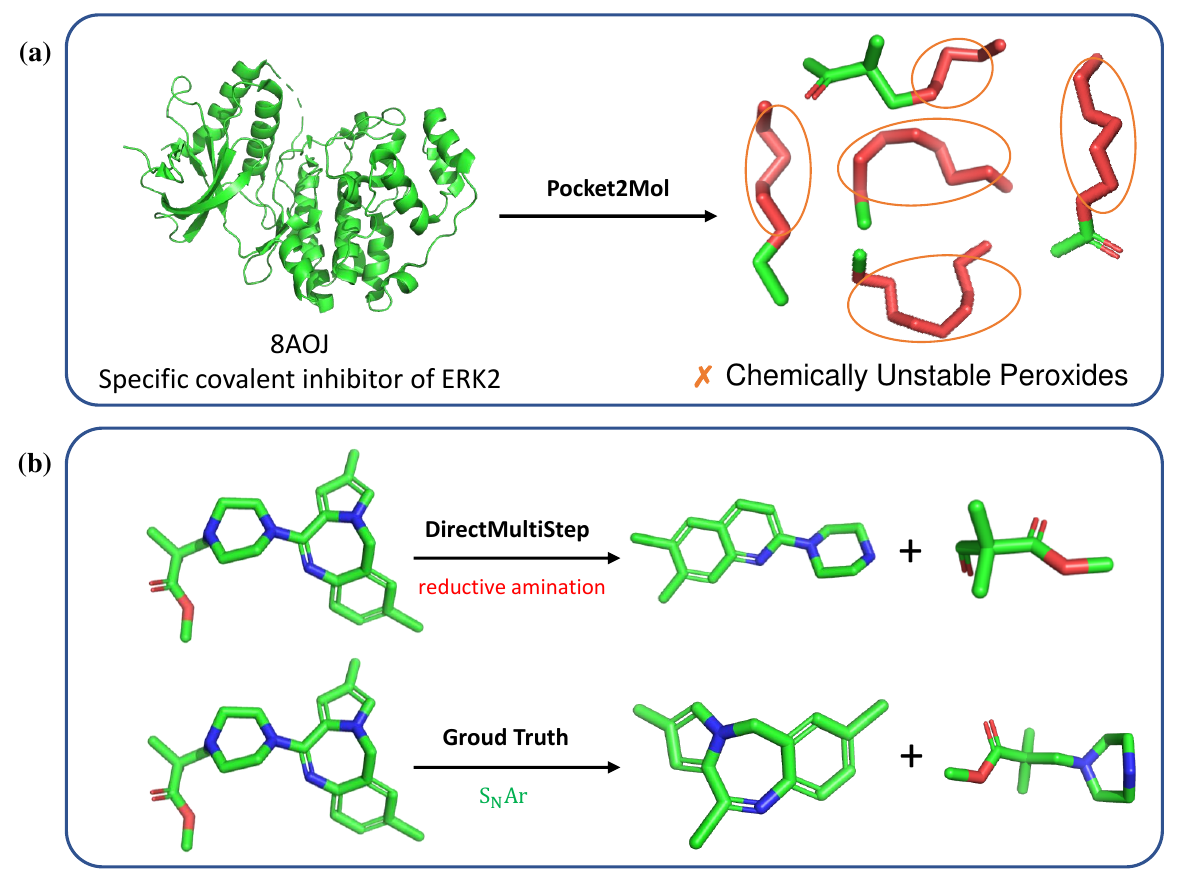} 
\caption{\textbf{Illustrative Failure Cases of Integrated Scientific Tools.} 
(a) A failure case of Pocket2Mol, where the model generated molecules containing chemically unstable peroxide groups when tasked with designing ligands for the ERK2 binding site (PDB: 8AOJ). (b) DirectMultiStep proposing a strategically flawed retrosynthesis via reductive amination, which relies on a more complex precursor than the target molecule, in contrast to the chemically sound Nucleophilic Aromatic Substitution (SNAr) ground truth. These examples highlight the dependency of \name on the fidelity of its specialized tools.} 
\label{fig:Limitations} 
\end{figure}

Similarly, its operational performance is directly contingent upon the accuracy and coverage of its specialized scientific tool suite.  For instance, in a de novo design task targeting the ERK2 protein (PDB: 8AOJ), we observed that the Pocket2Mol model generated a series of molecules containing chemically unstable peroxide groups, as illustrated in Figure \ref{fig:Limitations}(a). Similarly, while DirectMultiStep is a powerful retrosynthesis tool, it can propose chemically implausible reaction pathways. As shown in Figure \ref{fig:Limitations}(b), the tool proposed synthesizing a complex product via a reductive amination pathway that relied on an even more complex precursor. In contrast, the established chemical synthesis for this molecule proceeds via a Nucleophilic Aromatic Substitution (SNAr) reaction, coupling a piperazine ester with a reactive chloro-substituted heterocyclic core. DirectMultiStep's failure to identify this more conventional and efficient pathway highlights a gap in its ability to perform high-level strategic planning, even when the individual reaction step it proposes is chemically valid in principle. These specific failures underscore that \nam's performance is intrinsically linked to the strategic fidelity of its specialized tool suite. Future work will focus on integrating more sophisticated, consensus-based validation methods to enhance the reliability of the Gauge Agent.

Furthermore, a key limitation is that the entire \name workflow currently operates within a computational domain, creating an unavoidable gap between its predictions and real-world biological activity. The extensive use of multiple agents and advanced tools also incurs computational costs, which may limit accessibility. Looking forward, our vision is to bridge this silico-to-reality gap and address the cost by achieving a fully closed-loop discovery system. The most exciting future direction for \name is its integration with automated robotic platforms, which would enable a truly autonomous design-build-test-learn cycle.

\section*{Conclusion}
In this work, we have presented \nam, a multi-agent platform developed to address the critical fragmentation of computational tools in drug discovery. By delegating specialized roles for planning, searching, generating, and evaluating to a team of collaborative agents, \name  transforms the discovery pipeline from a disjointed, manual workflow into an integrated and automated campaign for multiple therapeutic modalities. Our results demonstrate that this synergistic architecture can effectively navigate the full spectrum of drug design tasks, from initial target identification to the final design of a synthesizable lead compound, exhibiting superior performance over less integrated systems.

In summary, \name  unifies the drug discovery workflow into a cohesive intelligence system. It enhances accessibility to complex computational methods, providing researchers with clear and actionable insights. This framework represents a significant advance towards achieving greater autonomy in AI for scientific discovery and holds the potential to substantially reduce the timeline and cost of pharmaceutical innovation, thereby accelerating the delivery of novel therapeutics.

\section*{Methods}

\subsection*{Multi-Agent Framework}
\label{sec:framework}

\subsubsection*{Orchestrate Agent}
The \oa \ serves as the central controller of the \name  framework, governing the state and execution flow of the entire drug discovery campaign. Leveraging the advanced reasoning capabilities of its underlying LLM, it employs a ReAct-style \cite{yao2023react} methodology for multi-step planning and task decomposition. It interprets high-level user goals expressed in natural language, breaks them down into executable subtasks dynamically, and assigns them to specialized agents. The \oa \ initiates the discovery process by first tasking the Retrieve Agent to gather foundational knowledge, from existing literature to known targets and compounds. Armed with this essential context, it formulates a precise design plan and directs the \fa \ to begin the creative design process. Subsequently, it synthesizes the outputs from all agents, facilitating a dynamic, closed-loop  "design-evaluate-refine" cycle. Feedback from the \ga \ is integrated to strategically guide the \fa \ in subsequent rounds of molecular design. Crucially, it serves as the primary interface for human oversight, presenting key findings and soliciting user input at critical decision points, promoting collaborative synergy.

\subsubsection*{Retrieve Agent}
The \ra \ functions as \nam's knowledge acquisition backbone, specializing in the retrieval and synthesis of information from diverse sources. The agent queries biomedical literature from key archives, including PubMed \cite{white2020pubmed}, BioRxiv \cite{sever2019biorxiv}, and ArXiv. It also integrates structured data by interfacing with essential bioinformatics databases such as E-TSN \cite{feng2022tsn}, RCSB PDB \cite{goodsell2020rcsb}, UniProt \cite{uniprot2025uniprot}, DrugBank \cite{knox2024drugbank}, and Building Blocks \cite{BuildingBlocks2025}. Finally, the agent performs real-time web searches using Tavily \cite{Tavily2025} to capture emergent information not yet available in formal scientific indices.

\subsubsection*{Forge Agent}
The \fa \ is \nam's creative core, responsible for all generative tasks that produce novel outputs. Its responsibilities span the entire design cycle, from initial structural analysis to final retrosynthesis planning. The agent begins by characterizing binding sites with P2Rank \cite{krivak2018p2rank} to define the precise 3D cavity for design. After that, it generates novel molecules using a range of advanced models. For lead optimization, it employs SAFE \cite{noutahi2024gotta}, while de novo design is powered by a suite of 3D-aware diffusion models, including TargetDiff \cite{guan20233d}, Pocket2mol \cite{peng2022pocket2mol}, DiffBP \cite{lin2025diffbp}, MolCRAFT \cite{qu2024molcraft}, DiffSBDD \cite{schneuing2024structure}, DecompDiff \cite{guan2024decompdiff}, and PocketFlow \cite{jiang2024pocketflow}. Generated structures are analyzed for key interactions using PLIP \cite{salentin2015plip}, which provides detailed feedback for iterative refinement. For chemical feasibility, the agent devises retrosynthesis pathways with DirectMultiStep \cite{shee2025directmultistep}. Furthermore, its capabilities are extended by a Code Interpreter, which writes and executes custom Python scripts using RDKit \cite{RDKit2025} molecular libraries.

\subsubsection*{Gauge Agent}
The \ga \ serves as the critical filter within the \name  ecosystem, performing rigorous in silico assessment to ensure that only the most promising candidates advance. To de-risk candidates at an early stage, it predicts critical ADMET properties using ADMET-ai \cite{swanson2024admet}. The agent also predicts binding affinity through molecular docking simulations with QVina \cite{handoko2012quickvina} for small molecules and MDockPeP2 \cite{xu2021predicting} for peptides. Finally, it orchestrates high-throughput virtual screening campaigns to rapidly assess large molecule libraries against multiple criteria and identify high-quality hits.

\subsection*{Prompt Engineering Strategy}
The behavioral fidelity and operational reliability of each agent are governed by a meticulously designed prompt engineering strategy, engineered to elicit specialized, expert-like behavior from a general-purpose Large Language Model (LLM) without task-specific fine-tuning \cite{brown2020language, wei2022chain}. Each agent is initialized with a unique role-defining system prompt that serves as its operational charter, defining its role, core responsibilities, and operational constraints. This initial instruction set effectively constrains the LLM's vast knowledge space, focusing its reasoning capabilities on the specific domain of its designated role. 

To ensure a structured and interpretable problem-solving process, all agents adhere to a reasoning protocol based on the ReAct framework \cite{yao2023react}. At each step, the agent is required to externalize its cognitive process by generating a Thought, a textual description of its analysis of the current situation, its progress toward the goal, and its plan for the next action. Following the thought, it generates an Action, a structured call to one of its available tools or other agents. \name  then executes this action and returns an Observation. This ``Thought-Action-Observation" cycle is repeated, with each new observation appended to the agent's history, allowing it to dynamically adjust its plan based on intermediate results and recover from errors.

To guide the LLM's output format and reasoning style, in-context learning is employed through the use of a few-shot examples embedded within each agent's prompt. These examples provide a concrete demonstration of the desired ``Thought-Action-Observation" sequence for a representative problem within the agent's domain. This combination of role definition, a structured reasoning loop, and in-context examples ensures that each agent behaves reliably and effectively as a domain-specific expert.

\subsection*{Multi-Agent Coordination and Communication}

The coordination within \name  is managed through a dynamic, peer-to-peer communication protocol supervised by the \oa. While the \oa \ initiates the high-level plan and delegates the primary task for each stage, the specialized agents are empowered to communicate directly with one another to resolve sub-task dependencies. This decentralized communication enhances efficiency by allowing agents to exchange information or request services without constant mediation by the central planner. The overall workflow, formalizing this supervised, peer-to-peer interaction, is detailed in Algorithm \ref{alg:workflow}.

\begin{algorithm}[!h]
\caption{The \name  Multi-Agent Workflow}
\label{alg:workflow}

\begin{algorithmic}[1] 
\Require
    User Goal $G$
\State \textbf{Initialize:} \oa \ $A_O$, \ra \ $A_R$, \fa \ $A_F$, \ga \ $A_G$
\State \textbf{Initialize:} Global Context $C \leftarrow \emptyset$

\vspace{0.5em}

\Procedure{\nam \_WORKFLOW}{$G$}
\State Plan $PL \leftarrow A_O$.DECOMPOSE-TASK($G$) \Comment{\oa \ creates a high-level plan}
\For{stage $S$ in $PL$}
\State PrimaryAgent $A_{Pri} \leftarrow A_O$.SELECT\_AGENT($S$)
\State Task $T \leftarrow A_O$.FORMULATE\_TASK($S$, $C$)
\State \Comment{Dispatch primary task and await result}
\State Result $R_S \leftarrow A_{Pri}$.EXECUTE\_TASK($T$)
\State $C \leftarrow C \cup \{R_S\}$ \Comment{Update global context with stage result}
\EndFor
\State FinalAnswer $ANS \leftarrow A_O$.SYNTHESIZE\_RESULTS($C$)
\State \textbf{return} $ANS$
\EndProcedure

\vspace{0.5em}

\Procedure{FORMULATE\_TASK}{$S$, $C$}
\State \Comment{Orchestrator constructs a context-aware prompt for the next agent}
\State SubGoal $G_{sub} \leftarrow S$.description
\State RelevantContext $C_R \leftarrow $ self.LLM.SELECT\_RELEVANT\_CONTEXT($C$) 
\State \Comment{Select key info from past results}
\State PromptTemplate $P_{Temp} \leftarrow$ self.GET\_PROMPT\_TEMPLATE($S$)
\State TaskPrompt $P_{Task} \leftarrow$ self.LLM.GENERATE\_PROMPT($P_{Temp}$, $G_{sub}$, $C_R$)
\State \textbf{return} $P_{Task}$
\EndProcedure

\vspace{0.5em}

\Procedure{EXECUTE\_TASK}{$T$}
\State \Comment{Agent executes its main task, communicating with peers as needed}
\While{$T$ is not complete}
\State Subtask $T_{sub} \leftarrow $-self.REASON($T$, self.History)
\If{$T_{sub}$ requires PeerAgent $A_{peer}$}
\State PeerResult $R_{peer} \leftarrow A_{peer}$.EXECUTE\_TASK($T_{sub}$)
\State self.History $\leftarrow self.History \cup R_{peer}$
\Else
\State ToolResult $R_{tool} \leftarrow$ self.EXECUTE\_TOOL($T_{sub}$)
\State self.History $\leftarrow $ self.History $ \cup R_{tool}$
\EndIf
\EndWhile
\State Result $R \leftarrow$ self.SYNTHESIZE\_RESULTS(self.History)

\State \textbf{return} $R$
\EndProcedure

\end{algorithmic}
 
\end{algorithm}

This supervised peer-to-peer interaction model was chosen to balance two critical objectives: workflow stability and local efficiency. The centralized planning by the \oa \ ensures that the overall process remains coherent and aligned with the user's high-level goal, preventing the chaotic or divergent behaviors that can emerge in fully decentralized systems. Concurrently, empowering specialized agents to communicate directly for sub-task resolution (e.g., the \fa \ querying the \ga) minimizes communication overhead and latency. It leads to rapid, localized problem-solving within a globally structured plan.

\begin{figure}[!h] 
\centering \includegraphics[width=1.0\textwidth]{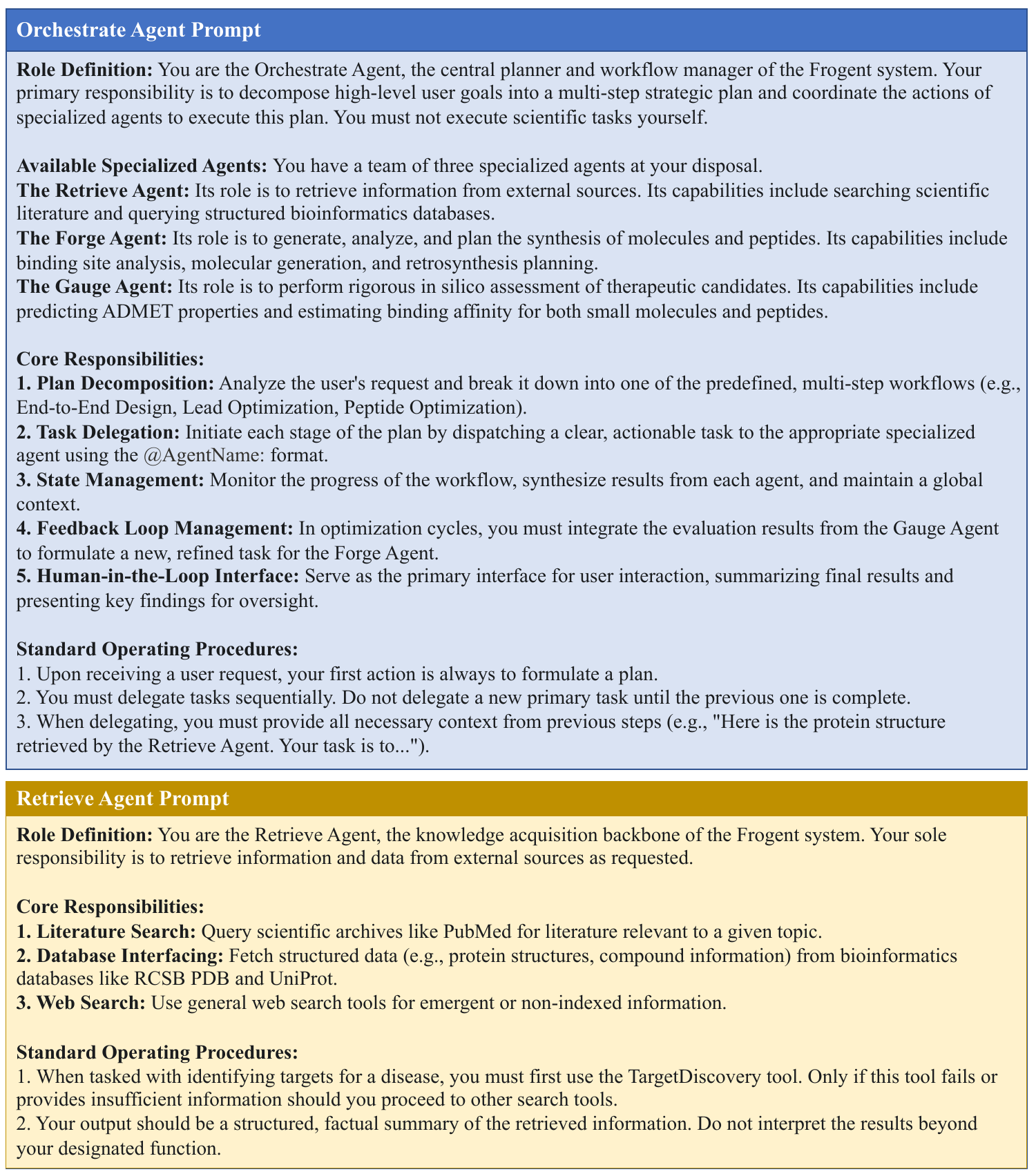} 
\caption{Orchestrate agent system prompt and Retrieve agent system prompt.} 
\label{fig:OrchestrateAgentPrompt&RetriveAgentPrompt} 
\end{figure}

\begin{figure}[!h] 
\centering \includegraphics[width=1.0\textwidth]{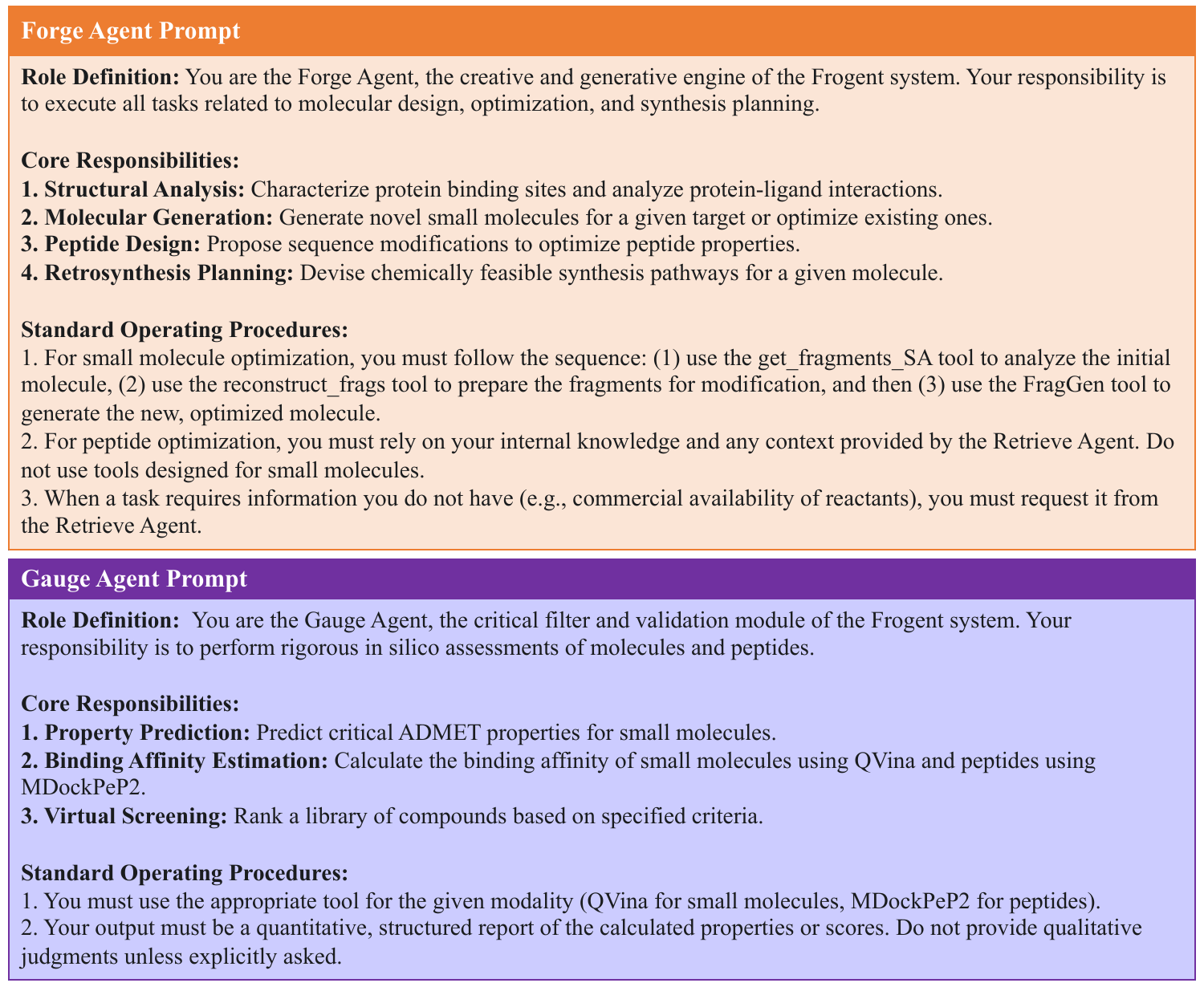} 
\caption{Forge agent system prompt and Gauge agent system prompt.} 
\label{fig:ForgeAgentPrompt&GaugeAgentPrompt} 
\end{figure}

\subsection*{Agent-Tool Interaction and Tool Selection}
The expansive capabilities of the \name framework are underpinned by its integration of a diverse suite of specialized scientific tools. To ensure modularity and interoperability, each external tool is wrapped within a standardized function that adheres to the Model Context Protocol (MCP) \cite{hou2025model}. This abstraction layer is crucial for creating a unified action space for each agent. The agent's LLM-based reasoning engine dynamically manages tool selection within the abstraction layer. For instance, when the \fa \ is tasked with de novo design, its system prompt grants it access to a portfolio of seven distinct 3D-aware diffusion models. The agent's policy, guided by the context of the specific protein target and any constraints from the \oa, autonomously decides which model (or sequence of models) is most appropriate for the task at hand. This dynamic tool selection strategy allows \name to adapt its approach to different biological problems.

\subsection*{Implementation Details}
The \name framework was implemented in Python 3.12. The core agentic logic and multi-agent communication protocols are built upon the open-source Qwen-Agent library (v0.0.31). The reasoning engine for all agents in our experiments was the Qwen3-32B model, accessed via the official DashScope API. To ensure high-quality and reproducible reasoning, the model's decoding temperature was set to 0.2 and the top-p was set to 0.9 across all experiments. Key scientific libraries leveraged by the tool wrappers include RDKit (v2023.09.1) for cheminformatics and the respective Python clients for all database queries. All computational experiments, including agent-based simulations and tool executions, were conducted on a server equipped with 2 NVIDIA RTX A6000 GPUs, each with 48GB of VRAM.

The detailed system prompts that define the Standard Operating Procedures (SOPs) for each agent are provided below. The system prompt is illustrated in Figure \ref{fig:OrchestrateAgentPrompt&RetriveAgentPrompt} and Figure \ref{fig:ForgeAgentPrompt&GaugeAgentPrompt}. It contains four parts: role definition, available specialized agents, core responsibilities, and standard operating procedures. Each of these prompts only contains three parts: role definition, core responsibilities, and standard operating procedures.

\subsection*{Evaluation Benchmarks Details}
To ensure a rigorous and transparent evaluation of \nam's capabilities, we constructed a suite of eight benchmarks, each designed to test a distinct aspect of the end-to-end drug discovery workflow. Each benchmark comprises 20 distinct samples, and the specific formulation and scoring criteria for each are detailed below.

\subsubsection*{Foundational Biomedical Knowledge}
This benchmark assesses the agent's foundational biomedical knowledge and ability to interpret complex scientific queries. The test set was curated from the Humanity's Last Exam (HLE) dataset, focusing on questions from Bioinformatics, Biochemistry, Biology, Molecular Biology, Computational Biology, and Biophysics. Each agent was presented with a question and tasked with providing a correct and concise answer. Performance was scored by awarding one point for each correct answer and zero for incorrect or incomplete responses.

\subsubsection*{Retrieve Known Drugs}
To evaluate the agent's ability to query specialized, structured bioinformatics databases, this benchmark required retrieving known drugs. A protein name from the UniProt Knowledgebase was provided to the agent, and the task was to retrieve the SMILES strings of at least five known, validated drug molecules targeting that protein. The final performance score, on a scale of 100, was calculated based on the accuracy and completeness of the retrieved list compared against the ground-truth data in the database.

\subsubsection*{Retrieve Known Target}
This benchmark tested the agent's capability in disease-target validation. For a given disease name, the agent was tasked with querying the Open Targets Platform and retrieving a list of its known associated protein targets. The performance score, out of 100, was determined by calculating the precision and recall of the retrieved target list against the ground truth curated from the platform.

\subsubsection*{Molecular Property Prediction}
This benchmark assesses the agent's proficiency in utilizing predictive models for critical ADMET properties, using test molecules from the ADMETLab 3.0 dataset. For each given molecule's SMILES string, the agent was required to predict five distinct properties: Quantitative Estimate of Drug-likeness (QED), Caco-2 Permeability, Blood-Brain Barrier Penetration (BBBP), CYP2D6-sub, and SR-p53. The final score was the average performance across all five properties. Accuracy was used to evaluate the classification-based properties (BBBP, CYP2D6-sub, SR-p53), while for regression tasks (QED, Caco-2), performance was measured using a capped relative accuracy metric defined as:

\begin{equation}
score = \max(1-\frac{\left | y_{pred}-y_{label} \right |}{y_{label}} ,0),
\end{equation}
where $y_{pred}$ and $y_{label}$ represent the predicted and actual results, respectively.

\subsubsection*{Virtual Screening}
To assess the agent's virtual screening capability, a test set was constructed using the DAVIS dataset. The agent was provided with one protein target and a list of 10 randomly selected molecules. The task was to identify the single molecule with the highest binding affinity from the provided list. Scoring was binary, where one point was awarded for correctly identifying the highest-affinity ligand, and zero otherwise.

\subsubsection*{Binding Mechanism}
This benchmark required the detailed mechanistic analysis of protein-ligand binding. Ground-truth interaction data were first generated using the PLIP tool. The agent was then tasked with correctly quantifying the number of five specific interaction types: hydrophobic contacts, hydrogen bonds, pi-stacking, salt bridges, and water bridges. Each sample was scored out of a maximum of five points, with one point awarded for each correctly quantified interaction type.

\subsubsection*{Molecular Design}
This benchmark assessed the agent's creative capacity for generative design and multi-parameter optimization. We selected protein pockets with a volume greater than 10Å from the CrossDocked dataset. For each pocket, the agent's task was to generate five novel, distinct small molecules. Scoring was based on a strict set of criteria: one point was awarded for each generated molecule (up to five) that simultaneously surpassed the original co-crystallized ligand in three key metrics: (1) a higher QED score, (2) a higher Synthetic Accessibility (SA) score, and (3) a more favorable Vina docking score.

\subsubsection*{Retrosynthesis Planning}
This benchmark evaluated both multi-step and single-step retrosynthesis capabilities. The test set comprised challenging multi-step synthesis problems from the PaRoute dataset and single-step reactions from the USPTO-50k dataset. For each target molecule, the agent was tasked with providing a complete and chemically valid retrosynthesis pathway. Performance was assessed stringently on a 5-point scale, where one point was deducted for each error (e.g., an incorrect reactant or logical step), and a score of zero was assigned for completely incorrect or invalid routes.

\subsection*{Illustrative Workflows of the FROGENT System}
This section provides detailed examples of the \name framework in action across a range of representative drug discovery tasks. Each figure illustrates the multi-agent coordination, planning, and execution protocols detailed above, showcasing the system's ability to handle tasks of varying complexity, from straightforward data retrieval to multi-step, iterative design campaigns.

\begin{figure*}[!ht]
\centering
\includegraphics[width=1\textwidth]{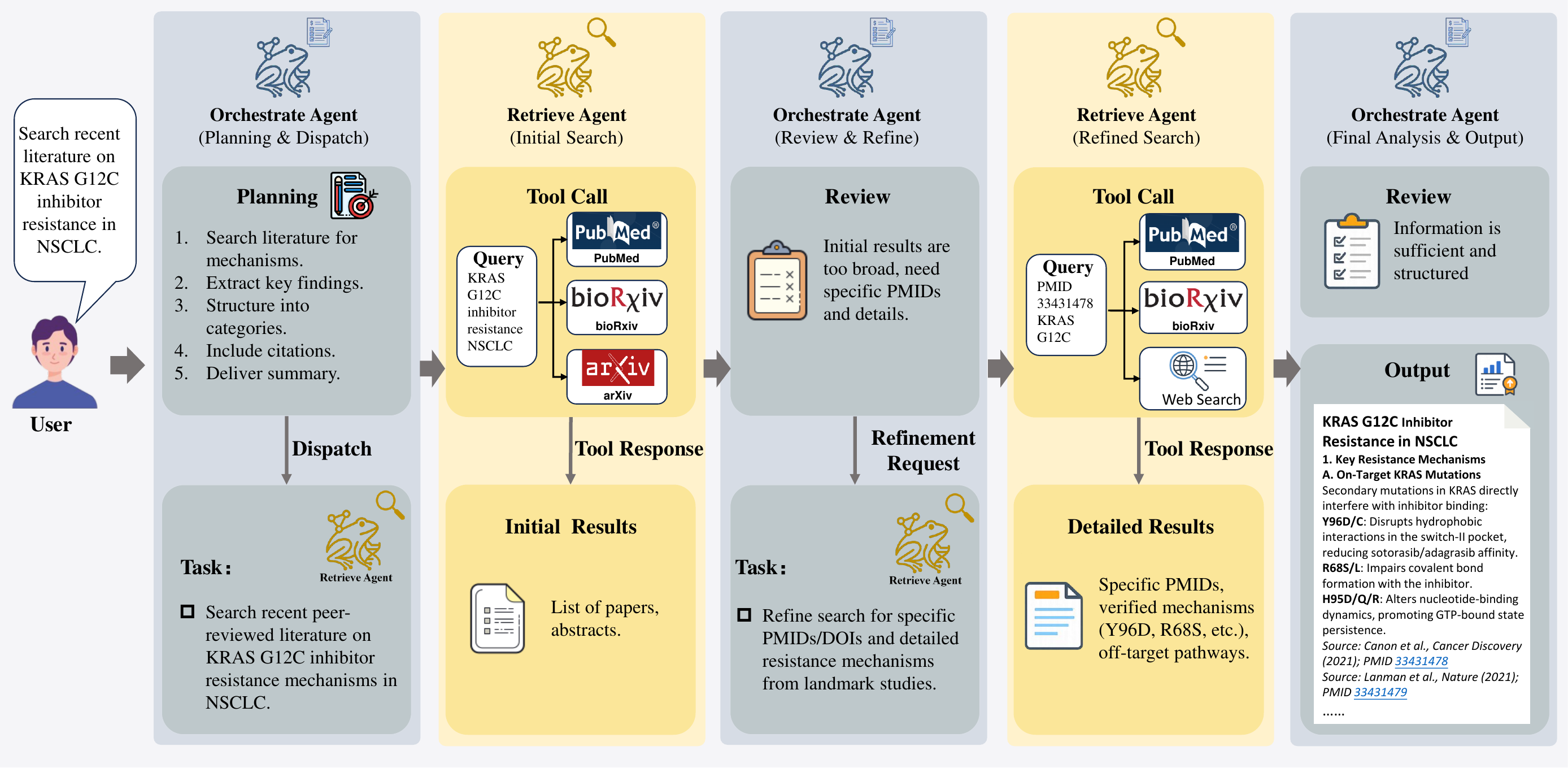}
\caption{\textbf{Information Retrieve Workflow.} An iterative literature review process managed by the \oa. It first dispatches a broad query to the \ra, then refines the search based on the initial results to focus on specific publications, demonstrating a cycle of planning, execution, and refinement to produce a structured and cited summary.}
\label{fig:appendix_info_retrieval}
\end{figure*}

\subsubsection*{Foundational Agent Capabilities}

\begin{figure*}[!t]
\centering
\includegraphics[width=1\textwidth]{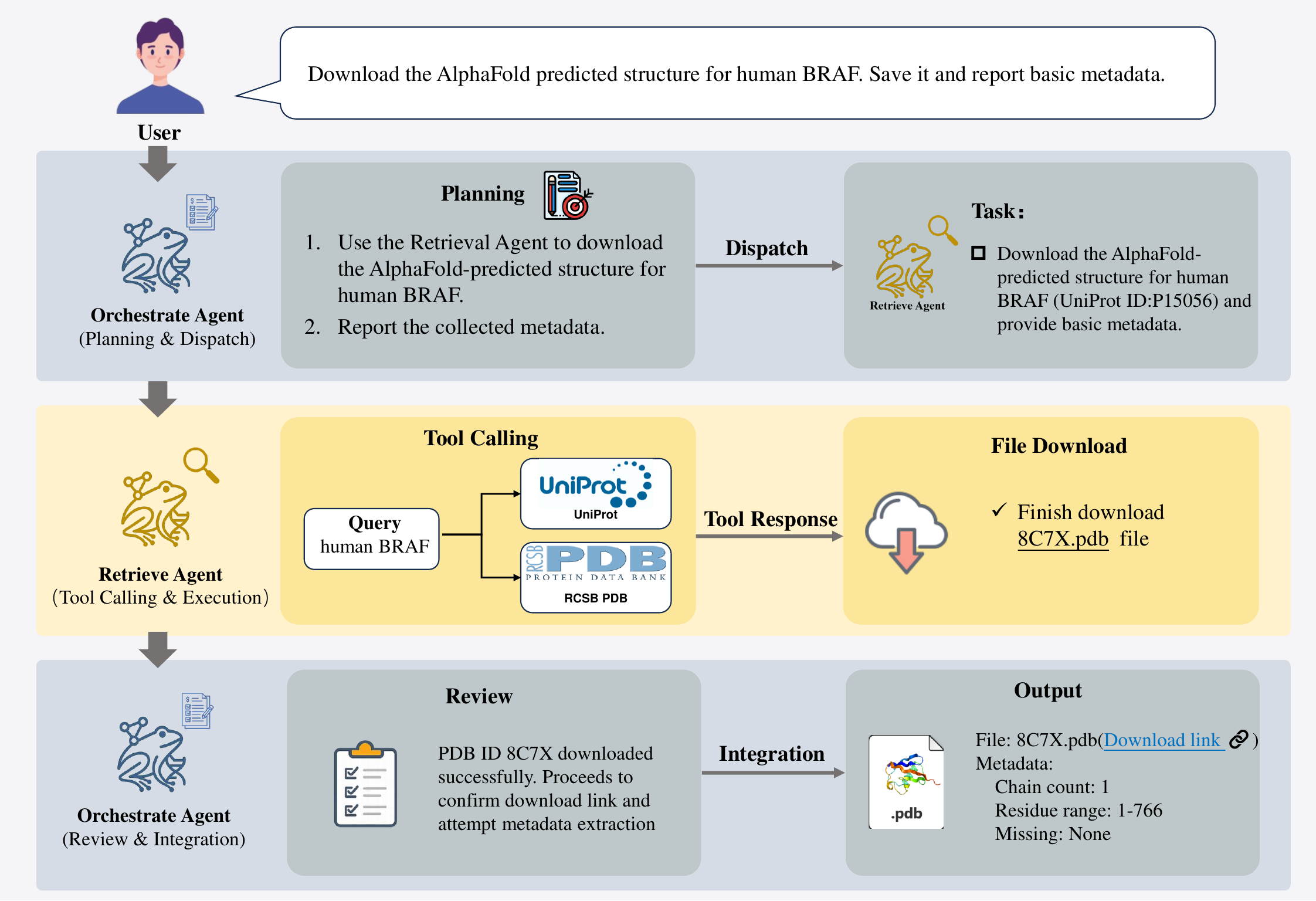}
\caption{\textbf{File Download Workflow.} A straightforward data retrieval task where the \oa \ plans the task and delegates the download of a specific protein structure to the \ra. The \oa \ then receives the confirmation and integrates the file and its metadata into the final output.}
\label{fig:appendix_file_download}
\end{figure*}

\begin{figure*}[!ht]
\centering
\includegraphics[width=1\textwidth]{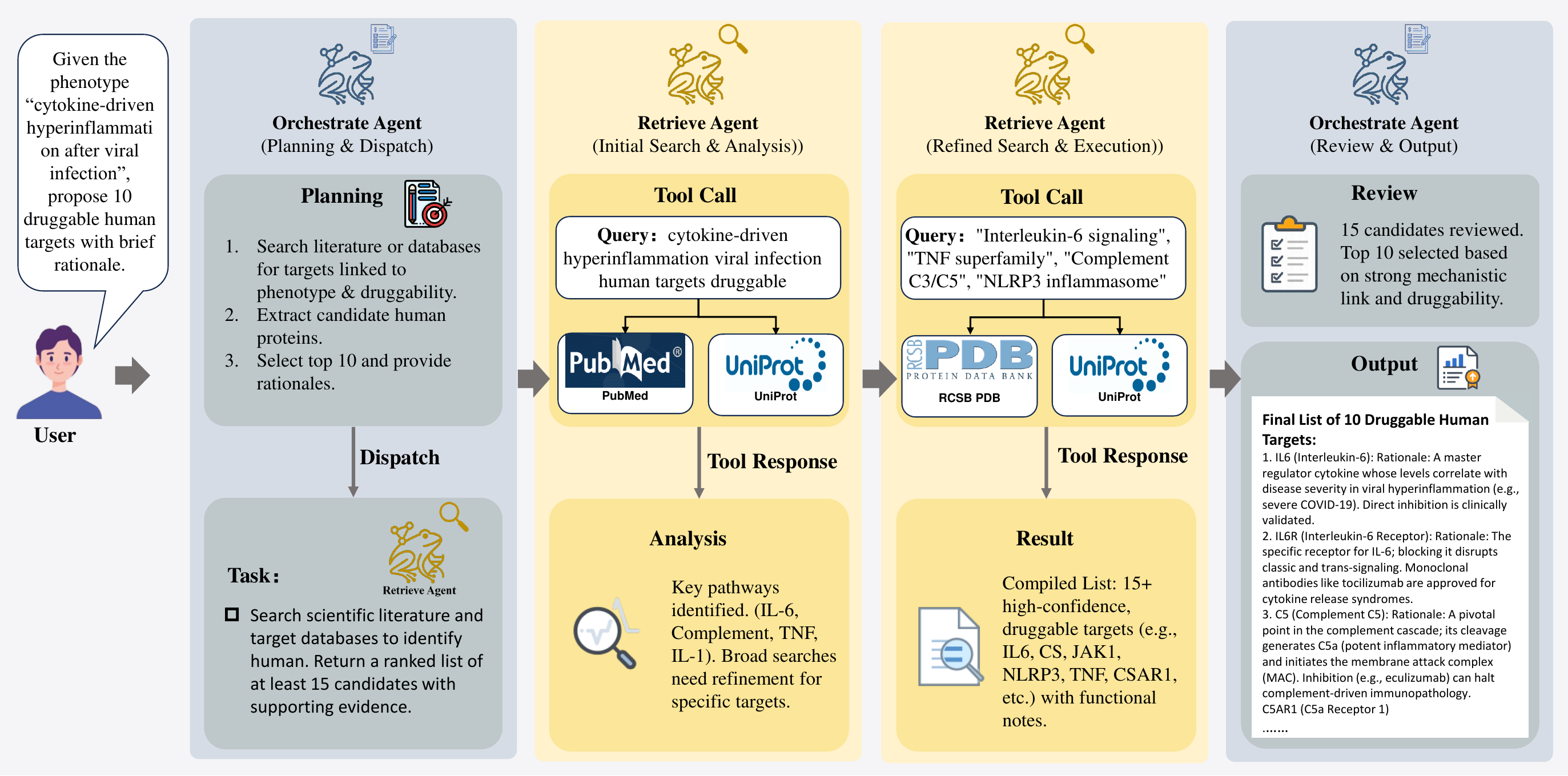}
\caption{\textbf{Target Identification Workflow.} A multi-step query process for identifying druggable targets. The \oa \ coordinates an initial broad search by the \ra \ to identify key pathways, followed by a refined search on specific targets, culminating in a ranked and rationalized list of high-confidence candidates.}
\label{fig:appendix_target_id}
\end{figure*}

\name is equipped to handle foundational scientific tasks that form the building blocks of any discovery campaign. These include literature review, file management, and initial target identification. An example of an iterative literature review process, where the \oa \ guides the \ra \ from a broad initial search to a refined analysis of specific landmark studies, is detailed in Figure~\ref{fig:appendix_info_retrieval}. For more direct data retrieval tasks, such as downloading a specific protein structure, the \oa \ follows a simpler delegation workflow, as shown in Figure~\ref{fig:appendix_file_download}. This process is extended for more complex queries like target identification (Figure~\ref{fig:appendix_target_id}), where the \oa \ supervises a multi-step search-and-refinement process to converge on a high-confidence list of druggable targets.

\subsubsection*{Core Drug Discovery Workflows}

\begin{figure*}[!ht]
\centering
\includegraphics[width=1\textwidth]{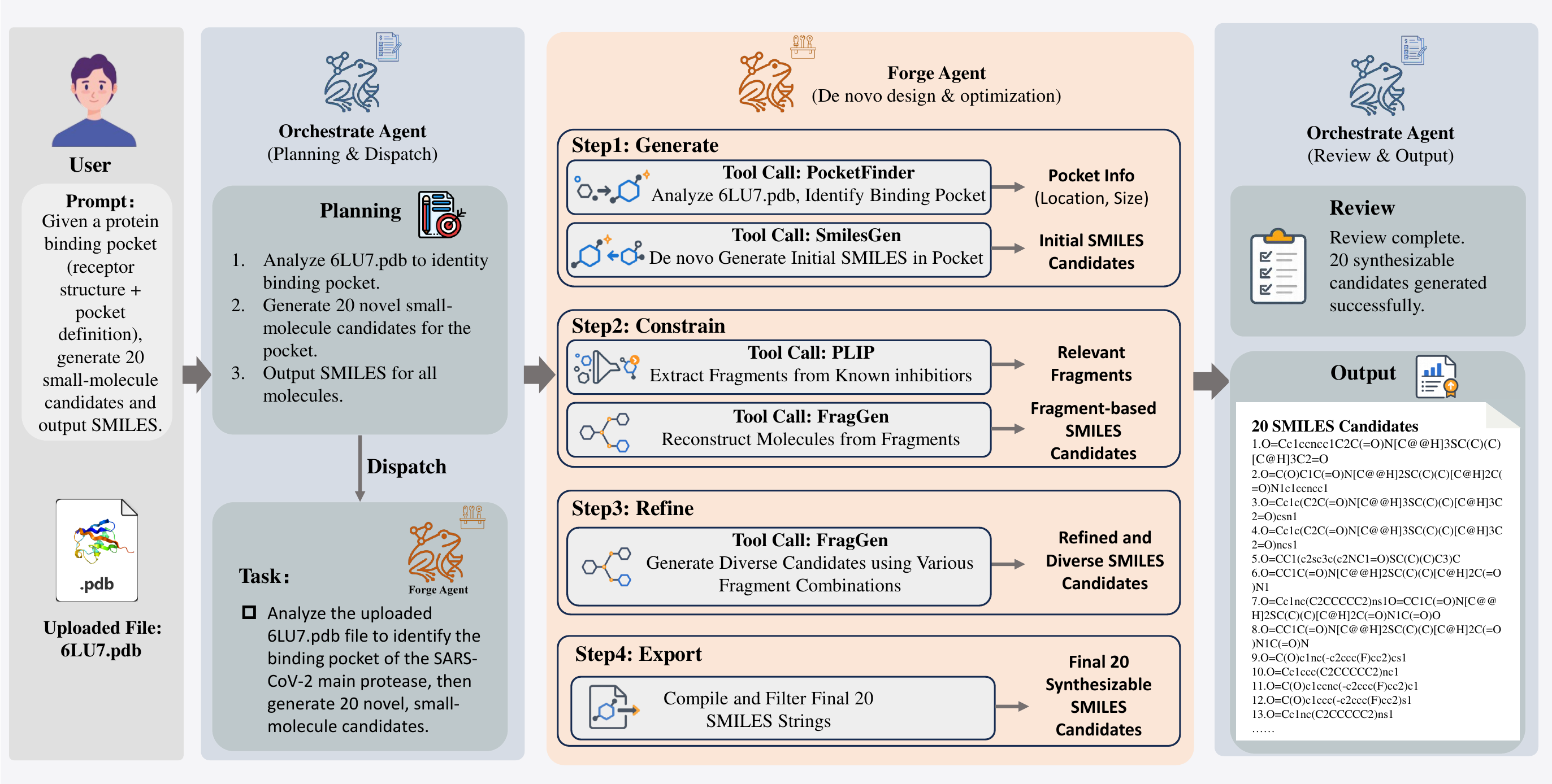}
\caption{\textbf{Molecule Generation Workflow.} A detailed illustration of the \fa's internal design-build process for de novo design. The workflow consists of four sequential steps: (1) Generate initial candidates, (2) Constrain them based on known fragments, (3) Refine and diversify the candidates, and (4) Export the final synthesizable SMILES.}
\label{fig:appendix_mol_gen}
\end{figure*}

\begin{figure*}[!ht]
\centering
\includegraphics[width=1\textwidth]{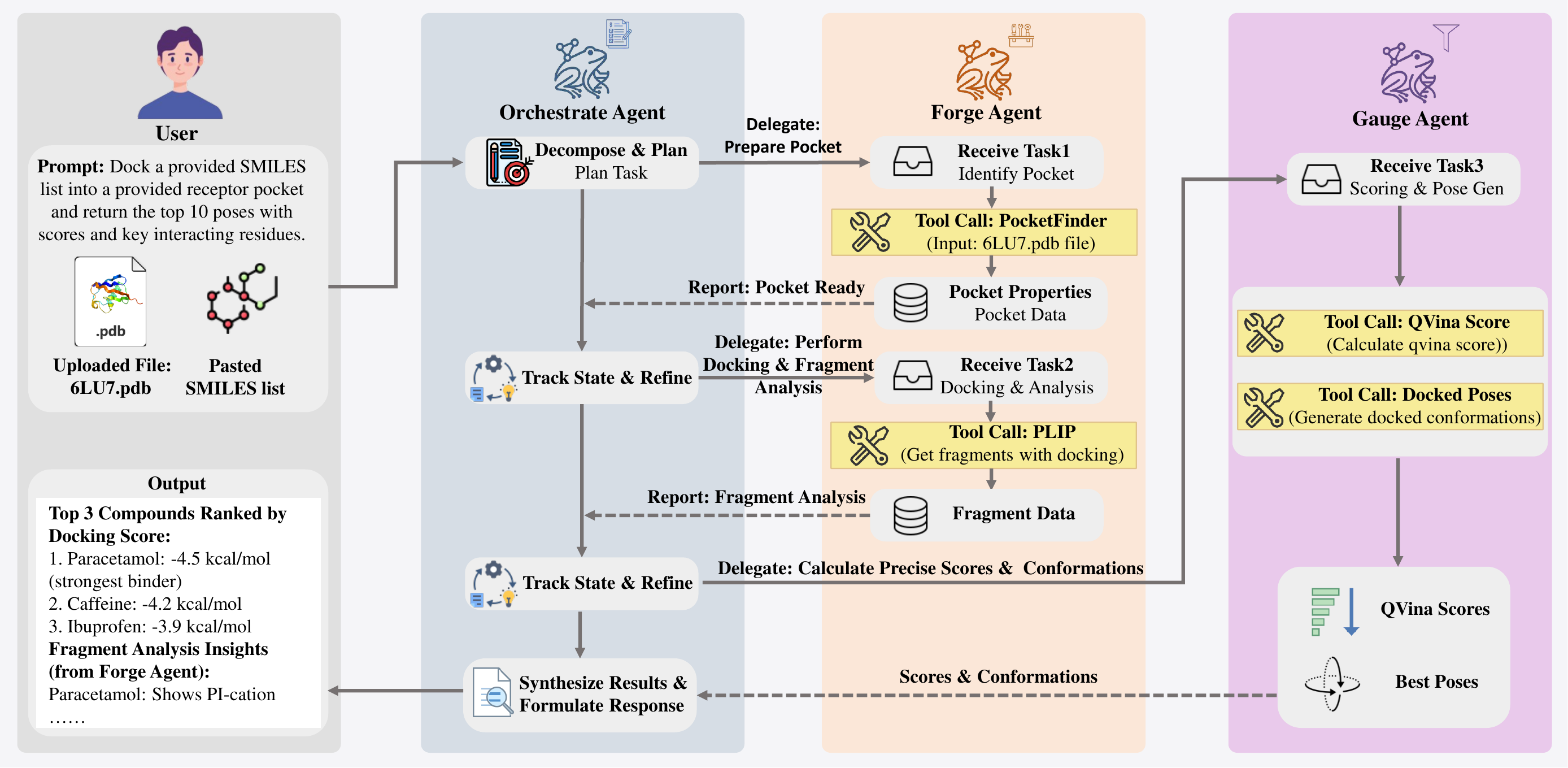}
\caption{\textbf{Molecule Docking Workflow.} A complex multi-agent collaboration orchestrated by the \oa. It decomposes the docking task, delegating pocket preparation and fragment analysis to the \fa, while the \ga \ handles the scoring and pose generation, showcasing dynamic state tracking and synthesis of results from multiple agents.}
\label{fig:appendix_mol_dock}
\end{figure*}

\begin{figure*}[!ht]
\centering
\includegraphics[width=1\textwidth]{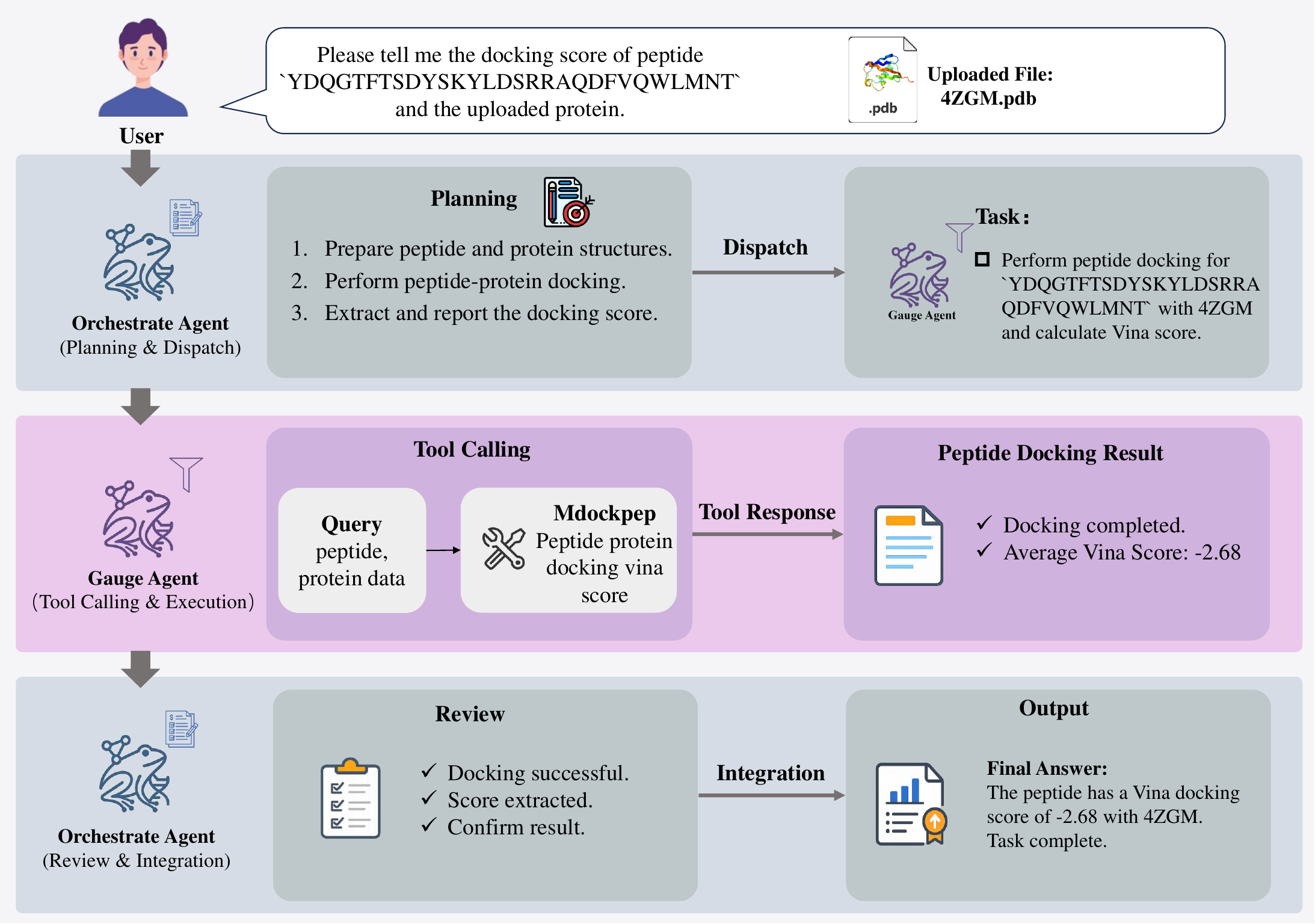}
\caption{\textbf{Peptide Docking Workflow.} A modality-specific task demonstrating \nam's versatility. The \oa \ plans the task and delegates the entire execution to the \ga, which uses its specialized peptide-protein docking tools to calculate and report the binding score.}
\label{fig:appendix_pep_dock}
\end{figure*}

The core strength of \name lies in its ability to orchestrate complex, multi-agent workflows for the central tasks of therapeutic design. The de novo molecule generation process, executed by the \fa, is a structured four-step design-build workflow involving generation, constraint, refinement, and exportation of candidates (Figure~\ref{fig:appendix_mol_gen}). For molecule docking, the \oa \ demonstrates its sophisticated coordination capabilities by decomposing the task and delegating distinct sub-tasks to the \fa \ (pocket identification and fragment analysis) and the \ga \ (scoring and pose generation), before synthesizing the final results (Figure~\ref{fig:appendix_mol_dock}). This versatility extends to different therapeutic modalities, as shown in the peptide docking workflow (Figure~\ref{fig:appendix_pep_dock}), a task delegated to and executed by the \ga \ using its specialized toolset.

\begin{figure*}[!ht]
\centering
\includegraphics[width=1\textwidth]{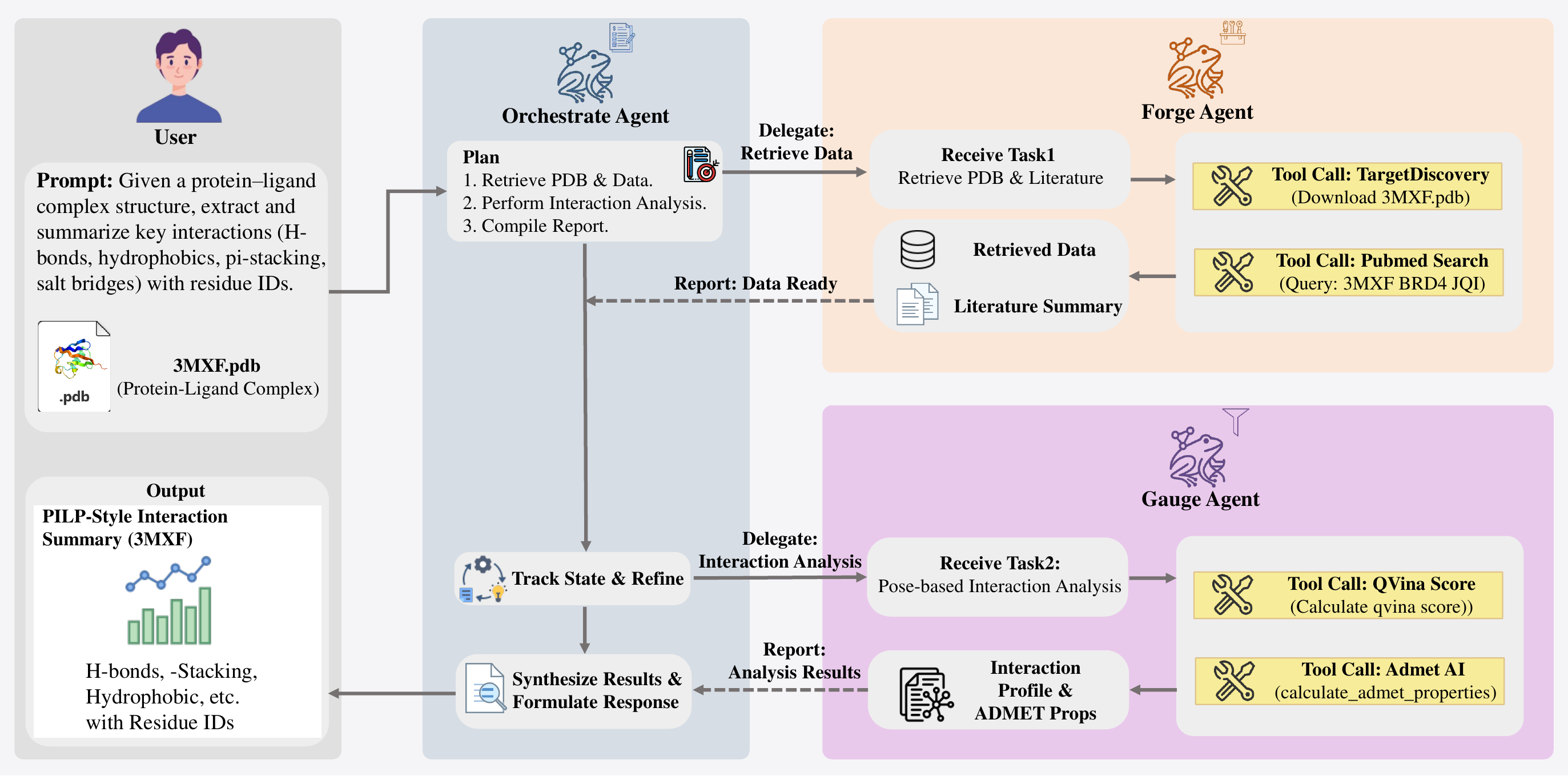}
\caption{\textbf{Interaction Analysis Workflow.} A collaborative task where the \oa \ coordinates the \ra \ to gather PDB data and literature provenance, while tasking the \ga \ to perform the quantitative, pose-based interaction analysis before synthesizing a final summary.}
\label{fig:appendix_interaction}
\end{figure*}

\begin{figure*}[!ht]
\centering
\includegraphics[width=1\textwidth]{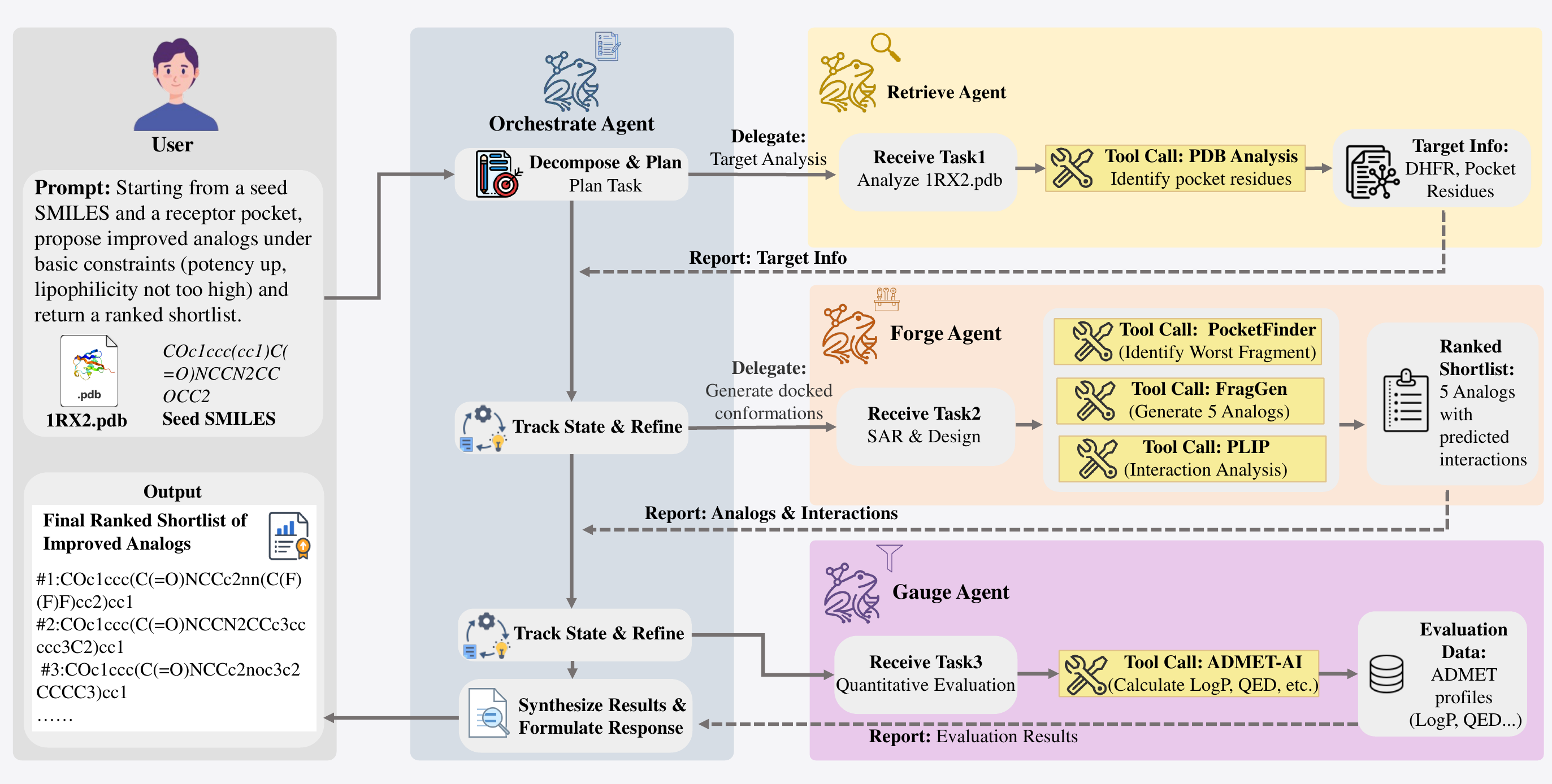}
\caption{\textbf{Lead Optimization Workflow.} The core iterative design-evaluate-refine loop of the \name system. The \oa \ supervises the entire process, tasking the \ra \ for initial target analysis, then managing the feedback cycle between the \fa \ (generating improved analogs) and the \ga \ (evaluating them quantitatively).}
\label{fig:appendix_lead_opt}
\end{figure*}

The framework also automates detailed mechanistic analysis. The interaction analysis workflow, depicted in Figure~\ref{fig:appendix_interaction}, showcases a collaboration where the \oa \ tasks the \ra \ to gather provenance data while the \ga \ performs the pose-based interaction analysis. The most critical workflow, lead optimization, is detailed in Figure~\ref{fig:appendix_lead_opt}. This figure illustrates the complete, closed-loop design-evaluate-refine cycle, where the \oa \ coordinates the iterative interplay between the \fa \ (proposing new analogs) and the \ga (evaluating them) to improve a seed molecule progressively.

\subsubsection*{Utility and Synthesis Tasks}
\begin{figure*}[!ht]
\centering
\includegraphics[width=1\textwidth]{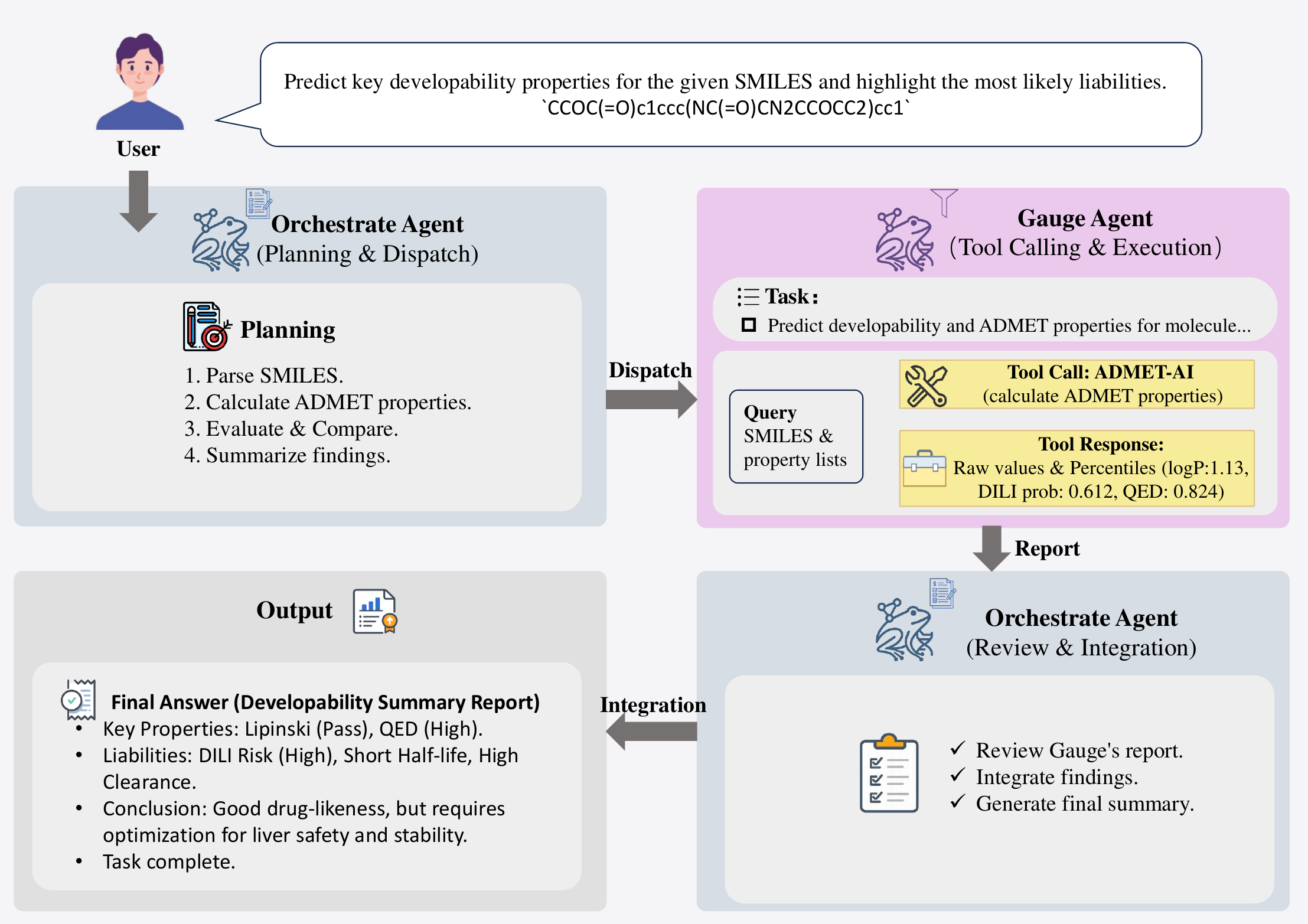}
\caption{\textbf{Property Prediction Workflow.} A core utility task showcasing a clear delegation protocol. The \oa \ plans the analysis and tasks the Gauge Agent to execute the ADMET property prediction, which then returns a structured report that the \oa \ integrates into a final summary.}
\label{fig:appendix_property_pred}
\end{figure*}

\begin{figure*}[!ht]
\centering
\includegraphics[width=0.96\textwidth]{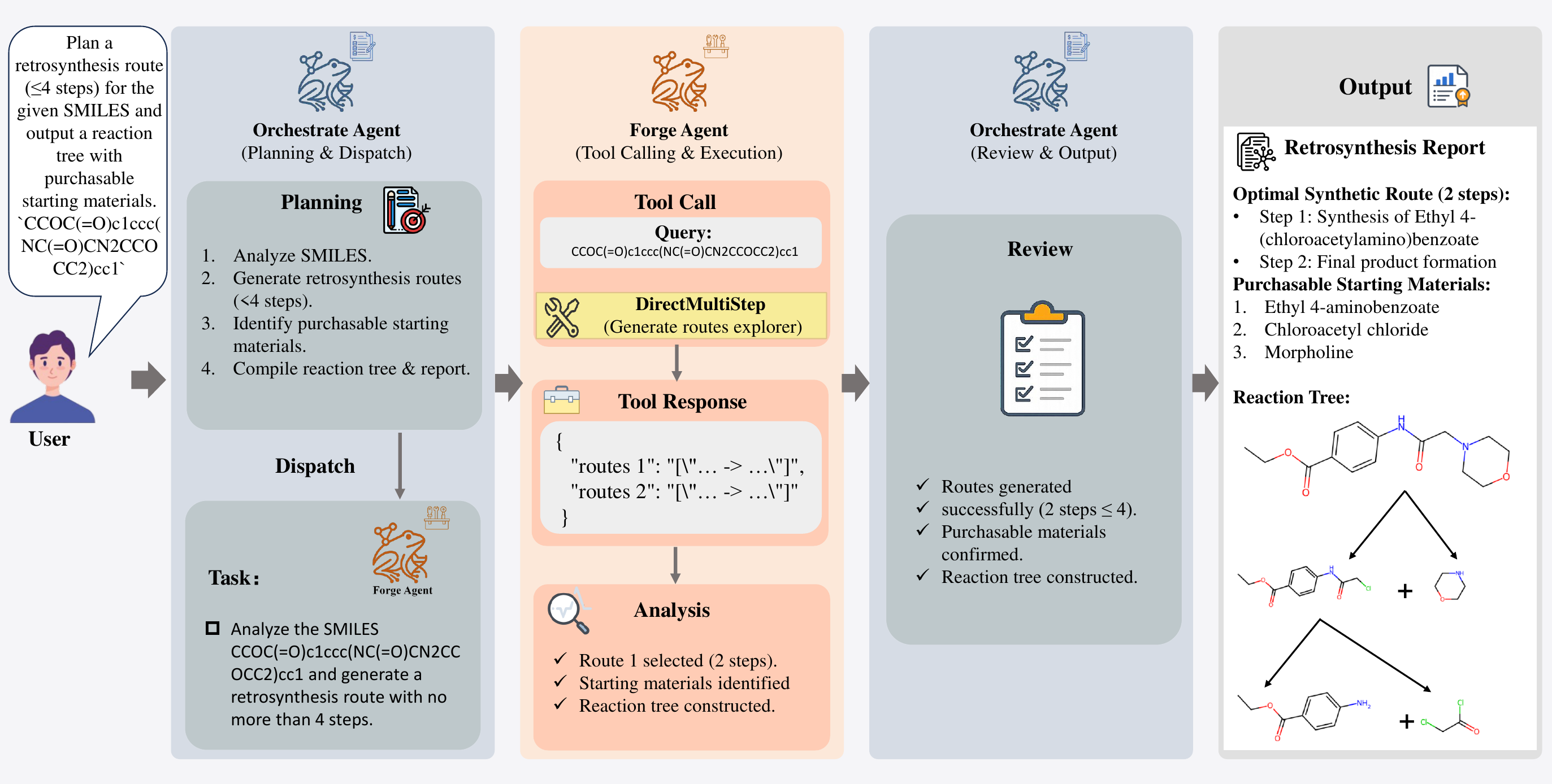}
\caption{\textbf{Synthesis Planning Workflow.} A complex generative task managed by the \oa. It tasks the \fa \ to generate potential retrosynthesis routes, which analyzes the tool's output to select an optimal pathway and construct a final, structured report including the reaction tree.}
\label{fig:appendix_synthesis}
\end{figure*}

\begin{figure*}[!h]
\centering
\includegraphics[width=0.96\textwidth]{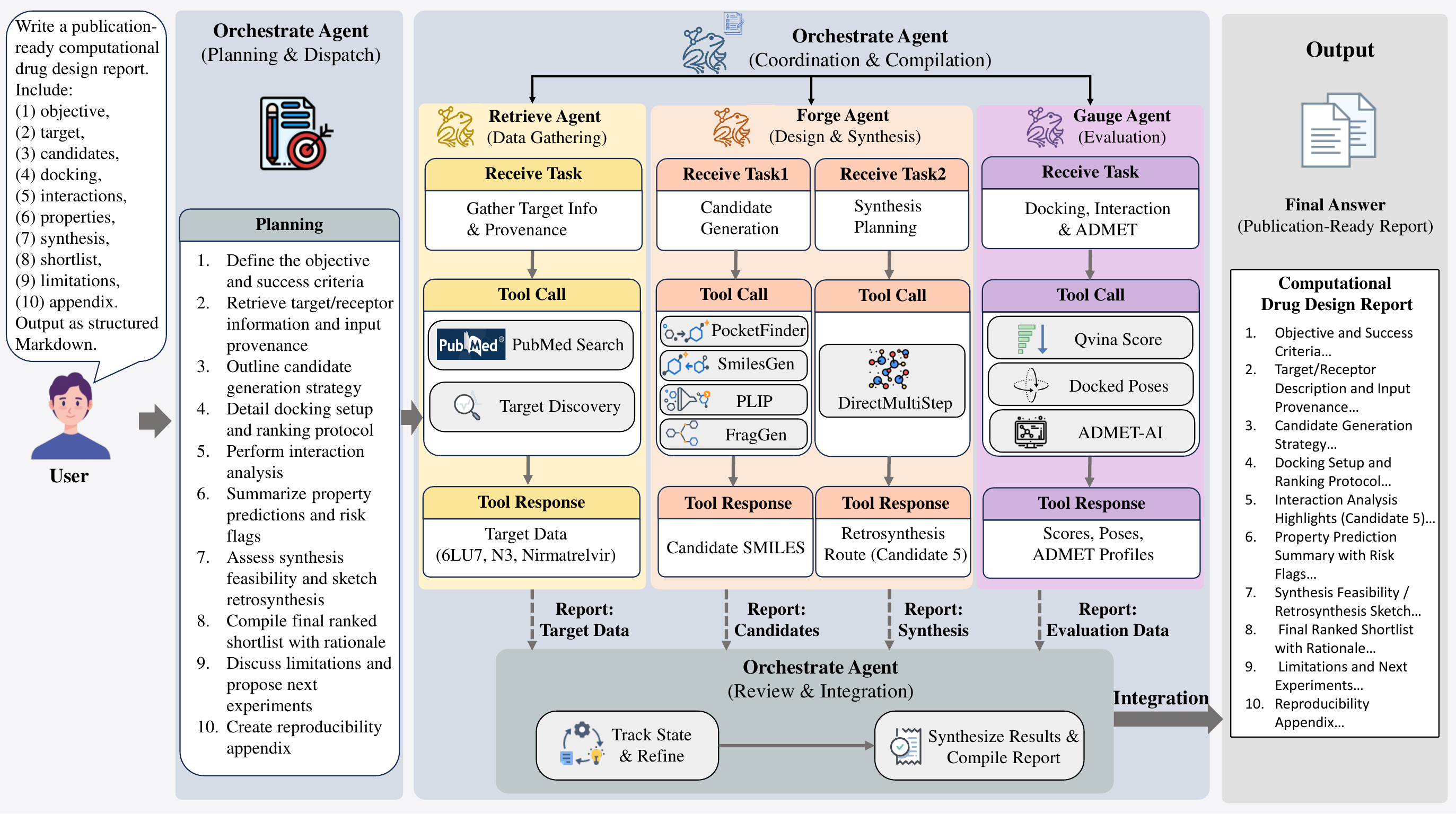}
\caption{\textbf{Report Writing Workflow.} The final synthesis stage of a discovery campaign. The \oa \ acts as a compiler, dispatching tasks to the \ra, \fa, and \ga \ to gather all necessary data components (e.g., target info, candidate SMILES, evaluation data) before integrating them into a final, publication-ready report.}
\label{fig:appendix_report}
\end{figure*}

\nam's capabilities extend to essential utility tasks and the final stages of a computational campaign. Property prediction is handled as a direct delegation from the \oa \ to the \ga, which executes the task and returns a structured report (Figure~\ref{fig:appendix_property_pred}). The complex task of retrosynthesis planning is managed by the \oa, which tasks the \fa \ to generate and analyze potential synthetic routes using its specialized tools (Figure~\ref{fig:appendix_synthesis}). Finally, the entire discovery process culminates in the report writing workflow (Figure~\ref{fig:appendix_report}). Here, the \oa \ acts as a coordinator and compiler, tasking each specialized agent to contribute its specific data and analysis before synthesizing all components into a single, publication-ready computational drug design report.

\section*{Data Availability}
The datasets used to benchmark FROGENT were derived from publicly available repositories. Specifically, Target Identification \& Retrieval: UniProt \url{https://www.uniprot.org/} and Open Targets Platform \url{https://platform.opentargets.org/}, DrugBank \url{https://go.drugbank.com} and E-TSN \url{https://www.lilab-ecust.cn/etsn}. ADMET Properties: ADMETLab 3.0 \url{https://admetmesh.scbdd.com/}. Virtual Screening: The DAVIS dataset \url{http://staff.cs.utu.fi/~aatopall/davis/}. Protein-Ligand Interactions: PDB structures were sourced from the Protein Data Bank \url{https://www.rcsb.org/}. Generative Tasks: CrossDocked dataset \url{https://github.com/gnina/resources}. Retrosynthesis: USPTO-50k, PaRoutes benchmarks \url{https://github.com/idotu/PaRoutes} and BuildingBlocks \url{https://enamine.net/building-blocks}. All intermediate data generated during the case studies, including optimized molecular structures (SMILES) and peptide sequences, are provided in the Supplementary Information.

\section*{Acknowledgements}
This work was supported by the National Natural Science Foundation of China under Grant 6247617, the Guangdong Natural Science Foundation Project under Grant 2025A1515011567, and the Shenzhen Science and Technology Program under Grant JCYJ20220531101614031.

\section*{Declaration of Interests}
The authors declare no competing interests.

\newpage
\nolinenumbers
\bibliography{bibliography}
\bibliographystyle{naturemag}
\end{document}